# Alloying to Tune the Bandgap of the *AM₂Pn₂* Zintl Compounds

Andrew Pike[1], Zhenkun Yuan[1], Muhammad Rubaiat Hasan[2], Smitakshi Goswami[1,3], Krishanu Samanta[4], Miguel I. Gonzalez[4], Jifeng Liu[1], Kirill Kovnir[2,5], Geoffroy Hautier[1,6,7]

1 Thayer School of Engineering, Dartmouth College, Hanover NH, 03755, USA

2 Department of Chemistry, Iowa State University, Ames, Iowa 50011, USA

3 Department of Physics, Dartmouth College, Hanover NH, 03755, USA

4 Department of Chemistry, Dartmouth College, Hanover NH, 03755, USA

5 Ames National Laboratory, U.S. Department of Energy, Ames, Iowa 50011, USA

6 Department of Materials Science and NanoEngineering, Rice University, Houston, TX 77005, USA

7 Rice Institute of Advanced Materials, Rice University, Houston, TX 77005, USA

Corresponding author email: geoffroy.hautier@rice.edu

## Abstract

The $AM_2Pn_2$ Zintl compounds are a large class of semiconductor materials that have a wide range of bandgaps and are mostly stable in the same crystal structure. Representative compounds $BaCd_2P_2$ and $CaZn_2P_2$ have recently been found to exhibit high visible light absorption and long carrier lifetime. Here we use high throughput first-principles calculations to study $AM_2Pn_2$ alloys for applications as tandem top cell absorbers (i.e., bandgaps around 1.8 eV) and far infrared detector materials (i.e., bandgaps lower than 0.5 eV). Using a first-principles computational screening workflow for assessing stability and electronic structure of alloys, we identify several promising candidates. These include $Ca(Cd_{0.8}Mg_{0.2})_2P_2$ with a suitable direct bandgap for use in tandem top cells on silicon bottom cells and $SrCd_2(Sb_{1-x}Bi_x)_2$ for far infrared detectors. We demonstrate that alloys of $AM_2Pn_2$ materials can be realized by experimentally synthesizing $Ca(Zn_{0.8}Mg_{0.2})2P_{2.}$

## Introduction

The $AM_2Pn_2$ family of Zintl compounds have recently emerged as a new class of semiconductor materials for optoelectronic applications. Among them, $BaCd_2P_2$ was identified computationally as a promising light absorbing material for thin film solar cells due to its suitable bandgap, high visible light absorption, and high "defect tolerance" (i.e. the absence of low-formation energy, deep carrier recombination centers), and confirmed experimentally to have strong band-to-band photoluminescence (PL) emission and a long carrier lifetime in powder samples[1,2]. $BaCd_2P_2$ nanoparticles have since been synthesized which also show long carrier lifetimes and bright PL emission, with the bandgap tunable by the quantum confinement effect[3]. These promising results also extend to $CaZn_2P_2$ from thin film synthesis, which is isostructural to $BaCd_2P_2$[4]. Even more recently, the photoelectrochemistry of $CaCd_2P_2$ has demonstrated activity for the oxygen evolution reaction under light with remarkable stability[5].

$BaCd_2P_2$ and $CaZn_2P_2$ represent just two members of a larger family of $AM_2Pn_2$ ($A$ = Ca, Sr, Ba, Mg, $M$ = Zn, Cd, Mg, and $Pn$ = N, P, As, Sb, Bi) materials, which have previously been reported as broadly stable, isostructural to $CaAl_2Si_2$ (space group $P\bar{3}m1$), and with a wide range of bandgaps from metals to beyond 2.5 eV[6]. They can be potentially used for a wide range of applications, such as solar cells and infrared detectors.

In this work, we explore applying the $AM_2Pn_2$ for tandem solar cells. The efficiency record for single crystal silicon solar cells has increased from 16.1% in 1980 to 25% in 1999 and then just an additional 1.1% to the current record of 26.1%[7]. They are in an area of diminishing returns wherein exponentially larger technological efforts will be required to achieve smaller amounts of efficiency increase because they are already approaching their theoretical efficiency limit. The tandem solar cell architecture is a way to boost the efficiency of a solar cell by incorporating a second junction to selectively absorb higher energy portions of the solar spectrum, leaving further room for efficiency gains[8,9]. By achieving higher efficiencies, the levelized cost of energy (LCOE) from photovoltaic (PV) systems can be reduced, since less material for the PV and module as well as less land for the installation are needed for a set amount of power generation[9,10].

We seek an appropriate tandem top cell solar absorber to pair with a Si bottom cell absorber to take advantage of its high efficiency, low cost, and long module lifetimes. Perovskites could be an attractive option for the top cell absorber due to their large bandgap and excellent low-temperature processability, but they suffer from rapid performance losses in operation[11]. The degradation rate effectively increases the LCOE as the module must be replaced more frequently[12]. Due to their promise as solar absorbers, the $AM_2Pn_2$ could also be attractive candidates for tandem top cell absorber layers.

Tandem top cells, as well as infrared detectors, require their bandgap to be tuned to absorb specific wavelengths of light. For instance, tandem top cell absorber materials require bandgaps higher than the optimal single junction bandgap (around 1.4 eV). It has been found that there are very few $AM_2Pn_2$ compounds with large and direct bandgaps[4,6]. $CaZn_2N_2$ and $CaCd_2P_2$ are the only two candidates, which have a direct bandgap of 1.9 and 1.62 eV, respectively. On the other hand, long wavelength infrared (IR) detector applications call for bandgaps between 0.1 and 0.5 eV[13].

Alloying is a widely used approach in the field of optoelectronics, where the bandgap of an alloy can be controlled by its composition. It is relatively unlikely that a material's bandgap precisely matches the optimum for a given application, so tuning it via alloying allows for more viable candidate materials. For example, thin-film and tandem solar cells rely on alloying to optimize the electronic quality of solar absorber materials, such as $CdTe_{1-x}Se_x$[14,15], $Cu(In_{1-x}Ga_x)Se_2$[16–18], $(Cu_{1-x}Ag_x)_2ZnSnS_4$[19,20], and the perovskite $BaZr(S_{1-x}Se_x)_3$[21,22]. Additionally, for IR detectors, bandgap tunability is one of the major strengths of $Hg_{1-x}Cd_xTe$[23,24].

Given their stability and isostructural nature, we can expect that many $AM_2Pn_2$ alloys could form. This provides an opportunity to design alloys with specific bandgaps. Several such alloys in this group have already been reported, although nearly all are antimonide and bismuthide compounds with small bandgaps for use as thermoelectric materials. Most notable is the $Mg_3Sb_{1-x}Bi_x$[25–27] alloy, as well as the related alloys $Mg_3((Sb_{0.5}Bi_{0.5})_{1-x}As_x)_2$[28] and $Mg_3(Bi_{1-x}P_x)_2$[27]. Other Mg-containing alloys are also experimentally reported, such as $Mg_{3-x}Zn_xSb_2$[29], $Mg_{3-x}Zn_xP_2$[30,31], $Ca(Zn_{1-x}Mg_x)_2Sb_2$[32], and $(Mg_{1-x}Ca_x)(Mg_{1-x}Zn_x)_2Sb_2$[33,34]. The nitride $Ca(Mg_{1-x}Zn_x)_2N_2$ has been synthesized and shown to have a large and tunable bandgap across the range of x from 0 to 1[35]. Finally, two $Ba(Zn_{1-x}Cd_x)_2Pn_2$ alloys have been synthesized: $Ba(Zn_{1-x}Cd_x)_2Sb_2$[36] with small amounts of Zn and $Ba(Zn_{1-x}Cd_x)_2P_2$[3] nanoparticles. These are notable due to the fact that $BaZn_2Sb_2$ and $BaZn_2P_2$ both form in the *Pnma* space group structures, unlike most other $AM_2Pn_2$.

Among the many possible $AM_2Pn_2$ alloys with the $CaAl_2Si_2$-type crystal structure, it remains to be identified which combinations will form specific, targeted bandgaps with direct bandgaps. Here, we employ a screening process based on first-principles calculations and search among all possible quaternary $AM_2Pn_2$ alloys for candidates that show favorable mixing and adequate bandgaps. We target the design of a material for the top junction of a tandem solar cell built with Si as the bottom cell. This approach takes advantage of mature Si PV technology, while further boosting the efficiency with a top cell. For such a device, the top cell ideally requires a 1.8 eV, direct bandgap for optimal efficiency[4,9]. We show that Mg alloying with Cd and Zn provides a promising pathway to achieve this, with $Ca(Cd_{1-x}Mg_x)_2P_2$ achieving these targets. To show that our computationally predicted alloys can be prepared experimentally, we demonstrate that $Ca(Zn_{1-x}Mg_x)$ alloys can be synthesized. We also explore alloys for use as long-wavelength IR detector materials and propose alloying on the anion site, such as $SrCd_2(Sb_{1-x}Bi_x)_2$.

Due to their lack of chemical long-range order, alloys are less straightforward to represent with first-principles calculations, and are therefore underexplored in high throughput studies. It is still uncommon to systematically screen semiconductor alloys. Yang *et al.* perform a high throughput screening for perovskite solar absorber materials, searching for alloys able to attain high efficiency in single junction solar PV devices[37]. They use a gradient of SQSs to compute properties along the continuous composition gradient. Other work includes that of Woods-Robinson *et al.* which introduced the "half space-hull" method to estimate stability and crystal structure of a very large set of potential alloys, but did not account for properties such as enthalpy of mixing[38]. Bhattacharya and Madsen explored metal-silicide alloys for thermoelectrics with a chemical mutation approach to identify promising alloying elements[39]. While they do not directly account for bowing of the bandgap, the effect of alloying is instead predicted by applying a volume deformation and then searching for element substitutions and compositions that can impose such a deformation. Here, we enumerate and exhaustively screen all possible quaternary alloys of the $AM_2Pn_2$ and explicitly evaluate the mixing properties, including enthalpy of mixing and bandgap bowing.

## Methods

### *Computational*

#### *DFT Input Parameters*

First-principles calculations were performed with the Vienna Ab initio Simulation Package (VASP) which uses the projector augmented wave (PAW) pseudopotential method[40–43]. For the screening process, calculations were automated with Atomate[44]. For structural relaxations we used the PBE functional[45] with a plane wave cut-off energy of 520 eV and an energy of $5\times10^{-5}$ eV/atom was used as the total energy convergence criterion for the self-consistent field (SCF) loop. Structural optimization was stopped with an energy convergence criterion of $5\text{x}10^{-4}$ eV/atom. The tetrahedral method with Blöchl corrections was used for Brillouin zone integration and the **k**-points were set with a density of 64 atoms/Å$^{-3}$. The relaxed structures were used to calculate the band structure with a **k**-point density of 20 per unit cell along each segment of the high symmetry **k**-point paths within the scheme of Setyawan and Curtarolo[46,47].

For thermodynamic quantities we used the r$^2$SCAN functional for an improved description[48–51]. A larger plane wave cutoff energy of 680 eV was used. For convergence criteria, the SCF loop was stopped when the total energy of an electronic step changed by less $5\times10^{-5}$ eV compared to the previous step and the structure optimization was stopped when the maximum residual atomic force was less than 0.02 eV/Å. The second order Methfessel-Paxton smearing scheme with a width of 0.2 eV was used. The smallest allowed distance between **k**-points was 0.22 Å$^{-3}$. The PBE_54 set of pseudopotentials was used.

To more accurately estimate the bandgaps of $Ca(Cd_{1-x}Mg_x)_2P_2$ alloys, hybrid functional calculations were performed using the parameterization of Heyd, Scuseria, and Ernzerhof (HSE)[52] with a Γ-only **k**-point grid. We used a mixing parameter of 0.25, a plane wave energy cutoff of 520 eV, and gaussian smearing with a width of 0.05 eV. HSE calculations were based on the PBE-relaxed structures.

*Alloy Property Interpolation*

To predict the bandgap of alloys along the entire compositional range, we apply a bowing parameter model:

$$E_g(x) = xE_{g,A} + (1-x)E_{g,B} - bx(1-x) \qquad \text{Eq. 1}$$

where $E_{g,A}$ and $E_{g,B}$ are the bandgaps of the unalloyed parent compounds and $b$ is the bowing parameter. Additionally, we define $\Delta E_g$ as:

$$\Delta E_g = E_g(x) - xE_{g,A} - (1-x)E_{g,B} \qquad \text{Eq. 2}$$

By finding the bandgap at $x = 0.5$ from the special quasirandom structure (SQS) calculation (see explanation below), we solve Eq. 1 to estimate $b$. Other works that estimate the bowing parameter from SQSs use several compositions and do a regression, but show the bowing is smooth across the compositional range leading us to believe a single SQS is reasonable to approximate the bowing parameter in the context of this investigation[53–55]. Additionally, Table S2 shows there is good agreement comparing known bowing parameters with those calculated as outlined above.

To assess stability along the composition range, we use the regular solution model:

$$H(x) = xH_A + (1-x)H_B + \Omega x(1-x) \qquad \text{Eq. 3}$$

And the enthalpy of mixing is given by:

$$\Delta H_{mix} = H(x) - xH_A - (1-x)H_B \quad \text{Eq. 4}$$

Where $H_A$ and $H_B$ are the enthalpies of the unalloyed parent compounds and $\Omega$ is the interaction parameter. Similarly to Eq. 1, this can be solved with the energy at $x = 0.5$ to determine $\Omega$.

The entropy of mixing is approximated with the ideal solution entropy of mixing

$$\Delta S_{mix} = -k_b a \sum_i x_i ln x_i \quad \text{Eq. 5}$$

Where, $k_b$ is the Boltzmann constant, $x_i$ is the mole fraction of each species on the sublattice with mixing and $a$ is a site factor for the sublattice where the mixing is occurring. For $AM_2Pn_2$ alloys, $a$ is 1/5 for mixing on the $A$ sublattice and 2/5 for the $M$ and $Pn$ sublattices.

*Screening process*

To search for promising candidate alloys, we perform a screening on all possible alloys in the $AM_2Pn_2$ family. For simplicity we restrict the search to quaternary alloys. Hypothetical alloys were based on the structure of the $BaCd_2P_2$ unit cell (space group $P\overline{3}m1$) with Ba, Cd, and P substituted for the necessary elements. For a preliminary screening of the alloys, a minimum-sized supercell was used to estimate the favorability of mixing in the alloy. Candidate compositions were substituted onto minimum-sized supercells for initial approximation of the phase stability of the alloy. For mixing on the $M$ and $Pn$ sites, this corresponds to a single unit cell, while for the $A$ site a 2×1×1 supercell is used because there is only one A site per unit cell. We note that this is an ordered approximation of a disordered alloy and emphasize that the information from this step is only used as a qualitative prescreening step to avoid further expensive SQS calculations of obviously unstable alloys. Alloys with $E_{hull} < 70\ meV/atom$ pass this step and move through the screening. This high threshold was chosen to due to account for the errors present from using the minimum-sized supercell as a coarse approximation of a disordered material.

To determine $E_{hull}$, the competing phases for each potential alloy were queried from the Materials Project and further relaxed with r$^2$SCAN[56,57]. Here, competing phases are those that bound the region of the phase diagram containing the alloy composition. The energy of the alloy is computed relative to the hull formed by these competing phases.

The compositions that pass the first screening step, are substituted onto an SQS for more refined estimates of stability and bandgaps[58,59]. One hundred twenty atom SQSs based on a 2×3×4 supercell were generated with mixing at the A, M, and Pn, sites separately which correspond to the compositions $A_{0.5}A'_{0.5}M_2Pn_2$, $A(M_{0.5}M'_{0.5})_2Pn_2$, and $AM_2(Pn_{0.5}Pn'_{0.5})_2$. From these calculations we get the bandgap with PBE and final total energy with r$^2$SCAN. SQSs were made using the mcsqs module in the Alloy Theoretic Automated Toolkit (ATAT) package[60,61]. Pair correlations up to 15 Å and triplet correlations up to 6 Å were considered in the objective function for the SQS generation.

To search for $AM_2Pn_2$ alloys with a specific bandgap, we solve for the composition where this is met by solving Eq. 1 with the bowing parameter we have found, and discard the alloy systems that are unable to reach this. We apply this bowing parameter with endmember bandgaps ($E_{g,A}$ and $E_{g,B}$) from HSE on PBE structures from our previous work[6]. This reduces the error associated with the systematic underestimation of the bandgap with PBE. We note that the bowing parameter is derived entirely from PBE, but due to a cancellation of errors, should be similar to the bowing parameter that would have been derived from HSE at a much lower computational cost[53,54,62,63]. This avoids HSE computations of the large SQSs and only requires HSE of the much smaller unit cell. See the SI for further justification of the error cancellation.

From the previous step, we also determine a composition, x, for each alloy system remaining. With this and Eq. 3, we estimate the $E_{hull}$ more precisely than in the minimum supercell step, to verify the alloy's stability. We use a stricter criterion of $E_{hull} < 20\ meV/atom$ for the candidates to keep.

***Experimental***

*Solid state synthesis*

*Warning: The amount of P in reaction ampoules should be kept to a minimum because at relatively high synthetic temperatures, the vapor pressure of P may be sufficient to cause the sealed ampoule to shatter or explode! Enclosing ampoules into secondary containment, such as ceramic beaker filled with sand or silica wool cocoon, and placing furnaces into well-ventilated space, such as fume hood, is highly recommended.*

The synthesis of the $Ca(Zn_{1-x}Mg_x)_2P_2$ crystals were carried out using Sn flux. The following high-purity elements were used for the reactions Ca (99.98%, Alfa Aesar), Zn (99.8%, Alfa Aesar), Mg (99.95%, Alfa Aesar), red phosphorus (98.90%, Alfa Aesar), and Sn (99.5%, Alfa Aesar). Ca, Zn, Mg and P were mixed in the stoichiometric ratios with 100 molar excess of flux: Ca:Zn:Mg:Sn=1:1.5:0.5:2:100 for $Ca(Zn_{1-x}Mg_x)_2P_2$ with $x = 0.25$. The mixtures, typically having a total mass of 8.5-9.5 g, were placed inside carbonized 9mm inner diameter silica ampoules inside an argon filled glovebox. A layer of silica wool was then placed on top of the reaction mixture such that it does not come into contact with the reactants. Chips of silica were then placed on top of the silica wool, and the ampoules were evacuated to <50 μTorr and sealed using a hydrogen-oxygen flame torch. After placing them in a muffle furnace in an upright position (silica chips at the top, reactants at the bottom), the ampoules were heated to 900°C, annealed at that temperature for 6 hours followed by slow cooling (4°C/hr) to 600°C, after which the ampoules were quickly taken out, inverted and rapidly centrifuged while hot. This allowed the molten Sn flux to be collected at the bottom of the ampoule along with the silica chips while the crystals either remained stuck to the walls of the ampoule or caught in the silica wool. The ampoules were cooled and then opened in ambient conditions. Grown crystals of $Ca(Zn_{1-x}Mg_x)_2P_2$ were generally smaller and thinner compared to grown $CaZn_2P_2$ crystals using a similar procedure. The same sample of Mg-alloyed $CaZn_2P_2$ was used for all subsequent analysis, except powder x-ray diffraction (PXRD).

Attempted synthesis of powdered phases $Ca(Cd_{1-x}Mg_x)_2P_2$ at x= 0.1, 0.2, and 0.3 were carried out using elemental Cd (99.95%, Alfa Aesar). without Sn flux, because the growth of undoped $CaCd_2P_2$ from Sn flux has not yet been achieved. The elements were weighed according to their stoichiometric ratios and loaded into a carbon-coated 9 mm inner diameter silica ampoule under an argon atmosphere in a glovebox. The ampoules were then evacuated to below 50 μTorr and sealed using a hydrogen-oxygen flame. The sealed ampoules were heated in a muffle furnace to 900°C over 9 hours, followed by a 48-hour annealing at that temperature. After cooling naturally in the turned-off furnace, the ampoules were opened inside an argon-filled glovebox. The resulting products were ground using agate mortars and used for further analysis.

*Scanning Electron Microscopy (SEM)*

Surface morphology and elemental composition of the Mg-alloyed $CaZn_2P_2$ samples were examined using a TESCAN Vega 3 scanning electron microscope equipped with an energy-dispersive X-ray spectroscopy (EDS) system. SEM imaging was conducted at an accelerating voltage of 30 kV under high-vacuum conditions. Crystal samples were mounted on aluminum stubs using carbon tape without any conductive coating, as the samples exhibited adequate surface conductivity to prevent charging during imaging. SEM was used to evaluate surface uniformity, grain morphology, and microstructural features in these samples. The SEM images were acquired at a magnification of 568× with a scale bar of 100 µm.

EDS analysis was performed using an EDAX detector operated through APEX software. Elemental mapping and point analyses were done at selected regions of the crystal surface to qualitatively and semi-quantitatively confirm the presence and spatial distribution of constituent elements (Ca, Mg, Zn, and P) and expected impurities (Sn, O, and C) in the Mg-alloyed $CaZn_2P_2$ sample.

*Single-Crystal X-ray Crystallographic Analysis*

Single-crystal X-ray diffraction data for pristine $CaZn_2P_2$ and Mg-alloyed $CaZn_2P_2$ were performed on three single crystals of each compound. Each crystal was coated with Paratone-N oil and mounted on MiTeGen loops. During the experiments, crystals were frozen at 100 K using an Oxford Cryosystems Cryostream 1000 low temperature system. Data were collected on a Rigaku XtaLAB Synergy-i diffractometer equipped with a HyPix-Bantam Hybrid Photon Counting Detector using Mo-Kα radiation ($\lambda$ = 0.71073 Å). Preliminary lattice parameters and orientation matrices were obtained from three sets of frames. Then the full data were collected using ω scans with a frame width of 0.5°. Raw data were integrated and corrected for Lorentz, polarization, and absorption effects using Rigaku CrysAlis$^{Pro}$ software[64]. Space group assignments were determined by examination of systematic absences, E-statistics, and successive refinement of the structures. Structures were solved by intrinsic phasing using SHELXT[65] and refined using SHELXL[66] as operated in the OLEX2[67] interface. No significant crystal decay was observed during data collection. Thermal parameters were refined anisotropically for all atoms. For each Mg-alloyed $CaZn_2P_2$ dataset, Zn and Mg atoms were constrained to have the same positions (EXYZ) and the same atomic displacement parameters (EADP). The total occupancy of the Mg/Zn site was set to be 100%. Refinement of these datasets without accounting for partial substitution of Zn by Mg gave rise to non-positive definite atomic displacement parameters and higher $R_1$ values. Additional crystallographic data are provided in Tables S5 and S6.

*Powder X-ray Diffraction*

Samples from the same synthesis batch as the sample mentioned above, were ground to a powder. Laboratory PXRD experiments carried out at room temperature were done using a Rigaku Miniflex 600 benchtop diffractometer having Cu-$K\alpha$ ($\lambda$ = 1.5406 Å) radiation and Ni-$K\beta$ filter. Patterns were collected at 40 kV tube voltage and 15mA tube current with 0.02° steps within a 3–90° 2θ range.

## Results

*Screening for Tandem Top Cell Solar Absorbers*

For a two-junction tandem solar cell to achieve maximum efficiency, the bandgap of the top and bottom cells must be selected according to the solar spectrum such that the current through them is matched. Taking advantage of Si as a mature PV technology, we search for a material whose bandgap is optimized to pair with it in a two-junction tandem solar cell. Due to Si's 1.12 eV bandgap, the top cell absorber should have a bandgap near 1.8 eV[4,9]. While there are a few $AM_2Pn_2$ with bandgaps close to this, most of them have indirect bandgaps, which would lead to poor optical absorption in a solar cell.

We perform a computational screening to search for materials for tandem top cell solar absorbers that have a bandgap of 1.8 eV and are thermodynamically stable. We screen all possible quaternary alloys in the $AM_2Pn_2$ ($A$=Ba, Sr, Ca, Mg; $M$=Zn, Cd, Mg, $Pn$=N, P, As, Sb, Bi) which is 240 possibilities. Figure 1 summarizes the steps in the screening. The screening begins by filling each candidate composition onto the unit cell of $BaCd_2P_2$. To accommodate the compositions where the $A$ site is shared by two elements, we use a 2×1×1 supercell since there is only one $A$ site per unit cell. The energy assessment at this stage is used as a go/no-go check for the stability. This is used as a qualitative way to rule out candidates which are unlikely to be miscible, despite being a relatively simple approximation of an alloy. Doing this removes about 1/3 of the 240 candidates, effectively reducing the search space for later steps. See Figure S1 for a comparison of the $E_{hull}$ from this approach and SQS.

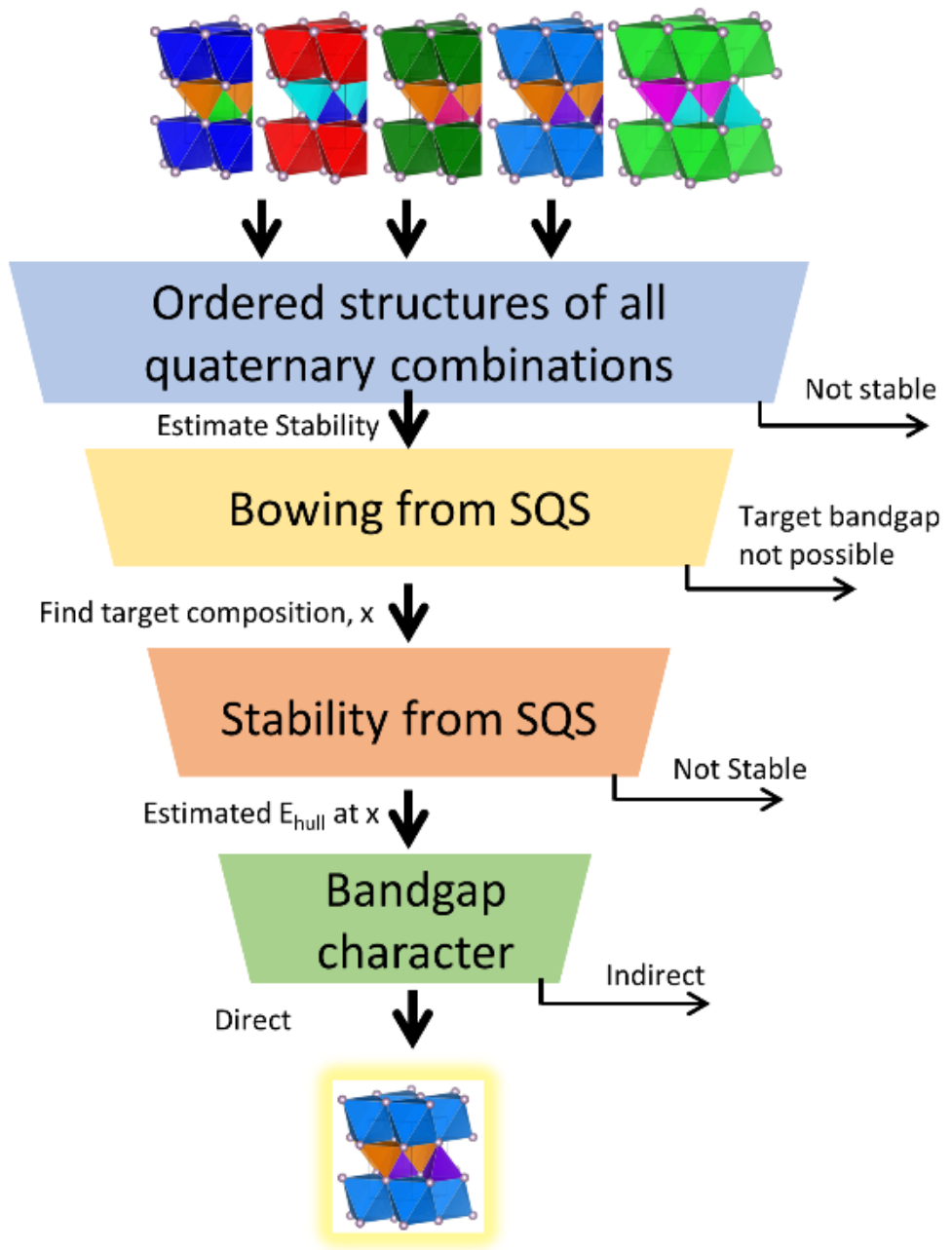

**Figure 1.** The screening process to search for tandem top cell solar absorbers.

Next, the bandgaps of the remaining candidates are computed from SQSs with PBE and used to estimate the bowing parameter for the alloy system. We overcome the underestimation of the bandgap from PBE by using this bowing parameter in conjunction with bandgaps from HSE on PBE relaxed structures for the endmembers (i.e., parent compounds). For tandem top cell solar absorbers, we target a bandgap of 1.8 eV. The alloy systems that are not able to achieve this bandgap anywhere in the compositions in $0 < x < 1$ are removed from the screening. Of those that remain, we estimate which $x$ will lead to a 1.8 eV bandgap.

From the original 240 hypothetical alloys, 26 candidate compositions are left at this point. We note that all of the remaining candidates contain Mg. Since Mg-containing $AM_2Pn_2$ compounds have the largest bandgaps compared to Zn- or Cd-containing ones, the addition of Mg is a rational way to increase the bandgaps of other compounds in the $AM_2Pn_2$ family.

From the SQS total energy and the composition ($x$) found in the previous step, the $E_{hull}$ can be found for the candidates. We apply a criterion of $E_{hull} < 20\ meV/atom$, which further removes 12 candidates, leaving 14. Table 1 shows the candidates remaining at this point in the screening. Since all of the candidates contain Mg, it seems the goal of designing an $AM_2Pn_2$ alloy for tandem top cell solar absorbers comes down to how well the other $AM_2Pn_2$ compounds can be mixed with Mg. Figure 2 summarizes the progress of each $AM_2Pn_2$ alloyed with Mg through the screening process. All interaction and bowing parameters found in the screening process are listed in the Table S4.

Here we examine the remaining candidates more closely. While it is expected that Mg would be present in many of the candidates, it has been suggested that high Mg content in compounds can lead to air and moisture-sensitivity[35,68]. Based on this concern, we remove $Sr(Mg_xCd_{1-x})_2N_2$, $Ba(Mg_xCd_{1-x})_2N_2$, $(Ba_xSr_{1-x})Mg_2P_2$, $Sr(Mg_xCd_{1-x})_2As_2$, $Ca(Mg_xCd_{1-x})_2As_2$, $Ca(Mg_xZn_{1-x})_2As_2$, $Sr(Mg_xZn_{1-x})_2As_2$, and $(Ba_xSr_{1-x})Mg_2As_2$ from further consideration since they would need to be relatively high in Mg content, as shown in Table 1. Regardless of air sensitivity, high Mg content would also likely lead to an indirect bandgap in phosphides and arsenides. Since both $CaMg_2P_2$ and $CaZn_2P_2$ have indirect bandgaps, it is likely that the band gap of $Ca(Mg_xZn_{1-x})_2P_2$ is also indirect for any $x$. $Sr(Mg_xZn_{1-x})_2P_2$ can be ruled out for the same reason.

This leaves $Ca(Mg_xZn_{1-x})_2N_2$, $Sr(Mg_xZn_{1-x})_2N_2$, $Ca(Mg_xCd_{1-x})_2P_2$ and $Sr(Mg_xCd_{1-x})_2P_2$ as the final candidates. For the two nitrides all parent compounds have direct bandgaps. From our calculations using HSE on PBE relaxed structures, $CaZn_2N_2$ already has a bandgap very close to the 1.8 eV target, so only a small amount of Mg is needed to increase the bandgap to the optimum. Previous experimental reports of the bandgap of $CaZn_2N_2$ show that our calculations are a slight underestimation of the bandgap and that it is about 1.9 eV[69,70], but it is still promising that the screening was able to recover this material and predict only a very small amount of magnesium is needed. $Ca(Mg_xZn_{1-x})_2N_2$ has previously been reported by Tsuji *et al.*[35] For the phosphides, pure $CaMg_2P_2$ and $SrMg_2P_2$ are not expected to be suitable candidates due to their indirect bandgaps and likely air sensitivity. However, the two phosphide candidate alloys are essentially $CaCd_2P_2$ or $SrCd_2P_2$ each with a small amount of Cd replaced with Mg, so they are expected to still have a direct bandgap, but it is increased in magnitude by alloying Mg. With such a small amount of Mg content we hope this avoids air and moisture sensitivity. In the following, we will focus more in-depth on the $Ca(Mg_xCd_{1-x})_2P_2$ alloy system, but we expect $Sr(Mg_xCd_{1-x})_2P_2$ to be similar.

Table 1 – Candidates from the screening process

| Formula | Composition (x) |
|---:|---:|
| $Ca(Mg_xZn_{1-x})_2N_2$ | 0.1 |
| $Sr(Mg_xZn_{1-x})_2N_2$ | 0.4 |
| $Sr(Mg_xCd_{1-x})_2N_2$ | 0.8 |
| $Ba(Mg_xCd_{1-x})_2N_2$ | 0.9 |
| $Ca(Mg_xCd_{1-x})P_2$ | 0.3 |
| $Ca(Mg_xZn_{1-x})_2P_2$ | 0.4 |
| $Sr(Mg_xZn_{1-x})_2P_2$ | 0.6 |
| $Sr(Mg_xCd_{1-x})_2P_2$ | 0.3 |
| $Ba_xSr_{1-x}Mg_2P_2$ | 0.6 |
| $Ca(Mg_xZn_{1-x})_2As_2$ | 0.8 |
| $Ca(Mg_xCd_{1-x})As_2$ | 0.8 |
| $Sr(Mg_xZn_{1-x})_2As_2$ | 0.9 |
| $Sr(Mg_xCd_{1-x})As_2$ | 0.9 |
| $Ba_xSr_{1-x}Mg_2As_2$ | 0.2 |

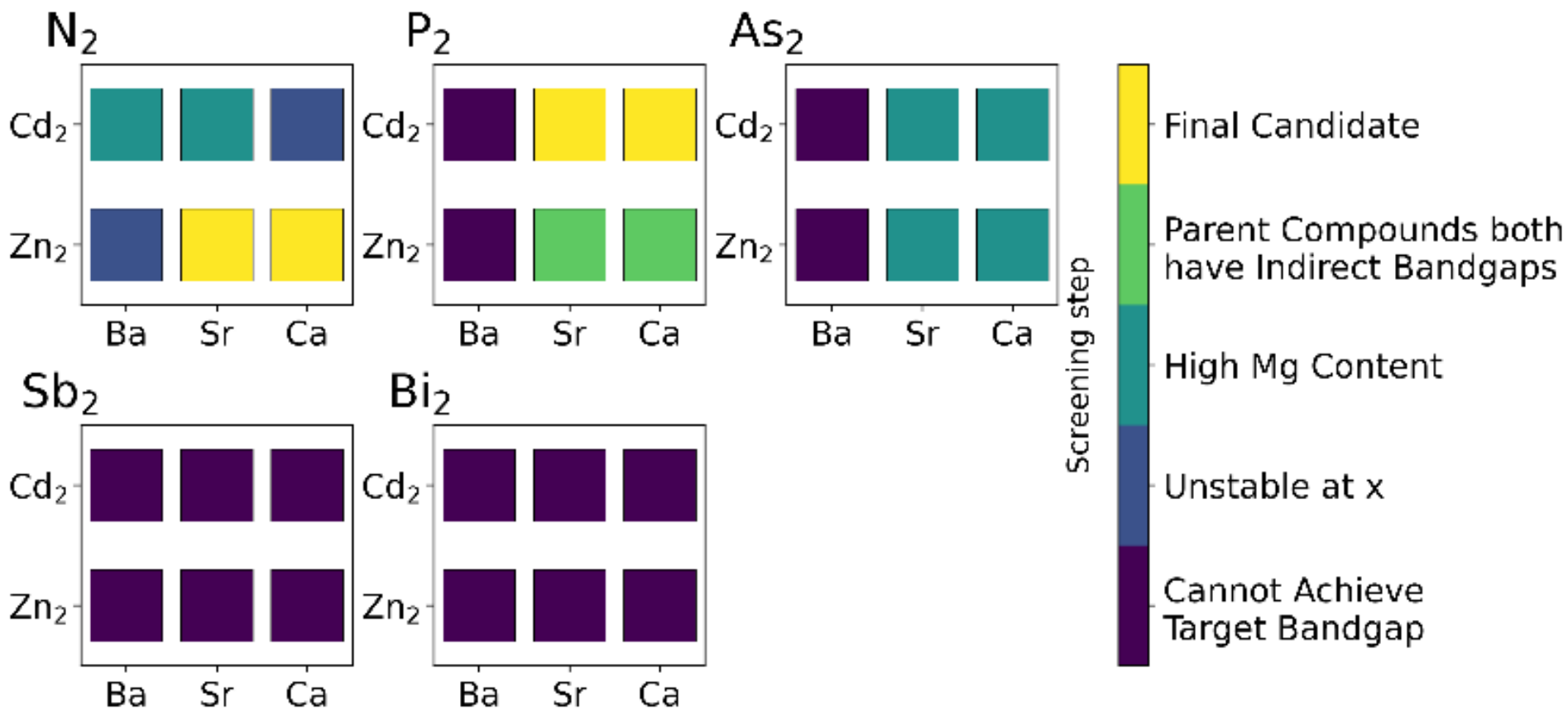

**Figure 2.** Screening of the *$AM_2Pn_2$* alloyed with Mg

*In-depth study of $Ca(Cd_{1-x}Mg_x)_2P_2$*

Here we investigate $Ca(Cd_{1-x}Mg_x)_2P_2$ from a series of SQSs to examine the entire compositional range. Figure 3 shows the mixing enthalpy and a quadratic fit as in the final term of Eq. 3. It can be seen that this system is well described by the regular solution model, and we derive an interaction parameter $\Omega = -24$ meV/atom. The negative mixing enthalpy across the entire composition range suggests that miscibility is predicted for this alloy system, that is, $Ca(Cd_{1-x}Mg_x)_2P_2$ will form a single phase without a miscibility gap. Including the effect of entropy at a reasonable synthesis temperature of 600°C through Eq. 5 further stabilizes the alloy phase, as shown in Figure 3. We note that the interaction parameter obtained from a single SQS computation in the previous step (−33 meV/atom) is similar to the one found here.

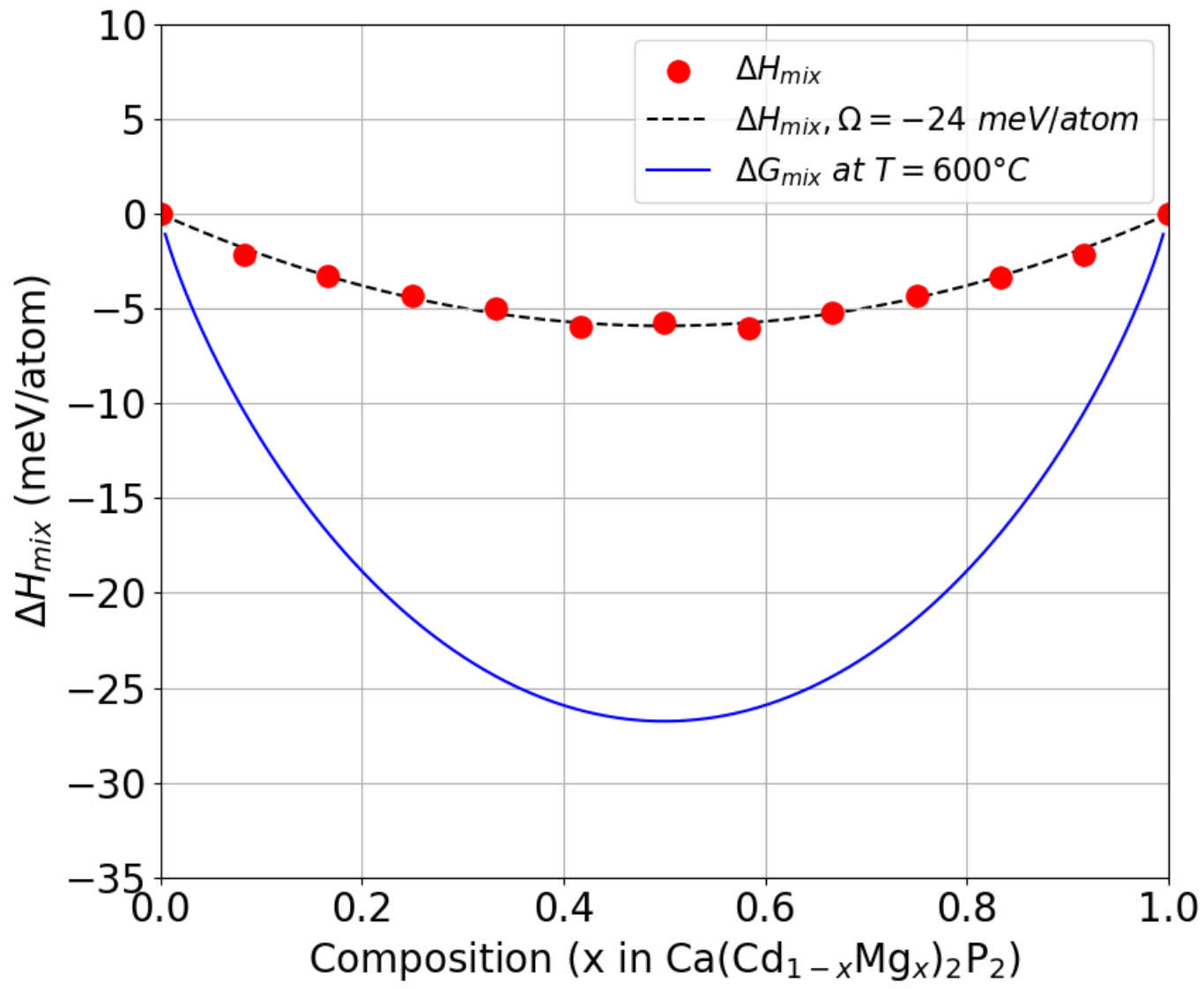

**Figure 3.** The mixing enthalpy and free energy of $Ca(Cd_{1-x}Mg_x)_2P_2$ from a composition gradient of SQSs. The red points are from the SQS calculations, the black dashed line is mixing enthalpy from the interaction parameter fit to the SQS calculations, and the blue line is the Gibbs' free energy including the effect of ideal mixing entropy.

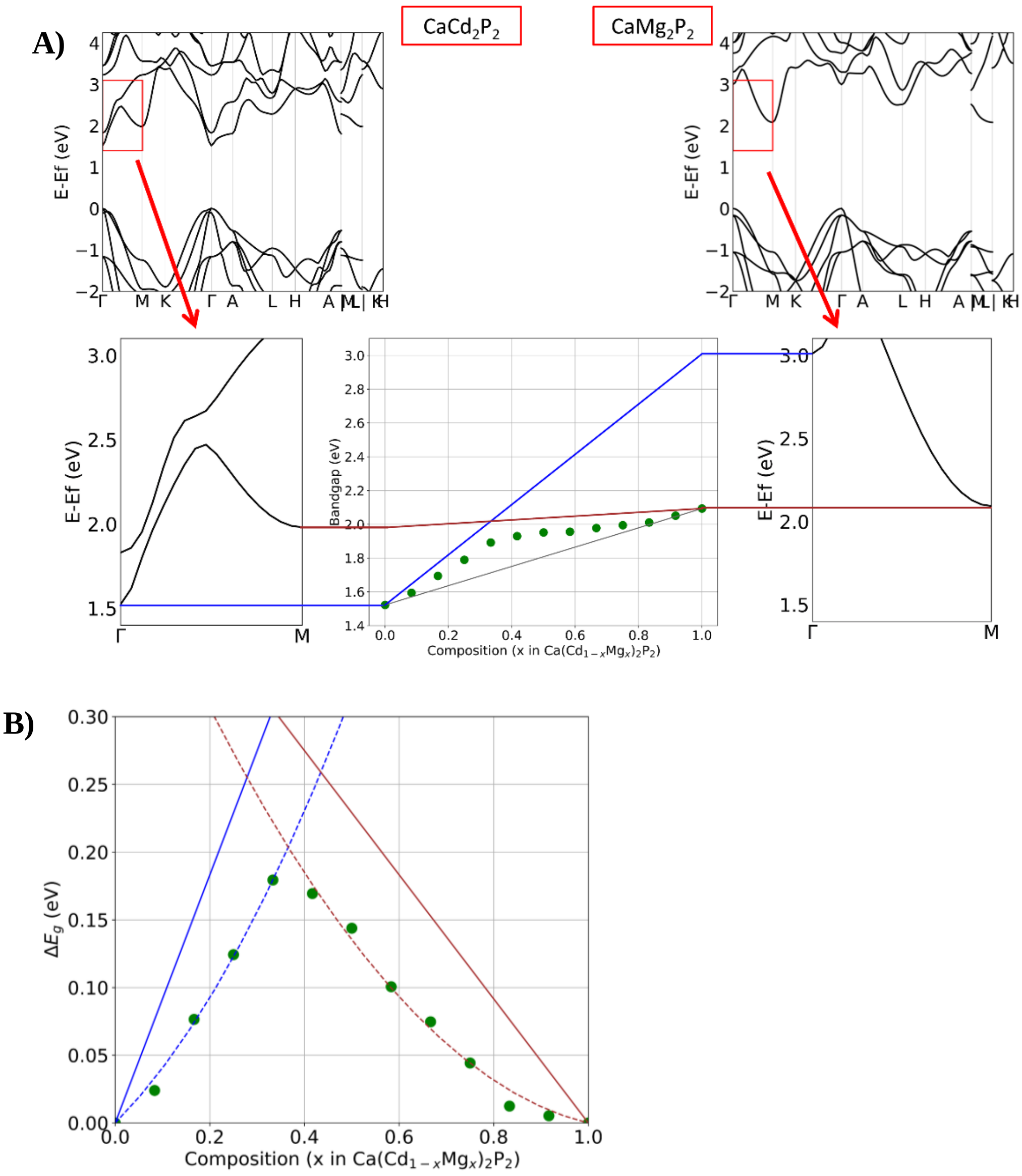

**Figure 4.** A) The band structures of $CaCd_2P_2$ and $CaMg_2P_2$. Here we focus on the **k**-point region from $\Gamma$ to $M$, at which the CBM of $CaCd_2P_2$ and $CaMg_2P_2$ are found, respectively. The lower outer plots zoom in on this region of the band structure. The lower central plot shows the change of the bandgap with composition. The blue line represents the linear change of the VBM at $\Gamma$ to the CBM at $\Gamma$ transition with composition and the brown line represents that of the VBM at $\Gamma$ to the CBM at $M$. The black line represents the weighted average of the bandgap of the $CaCd_2P_2$ and $CaMg_2P_2$. The green points are the calculated bandgaps. All plots are calculated with PBE and adjusted to HSE values with a rigid shift of the conduction band. B) $\Delta E_g$ versus composition derived from Eq. 2. The dashed blue line is the fit of the data with a bowing parameter to the direct bandgap from the VBM at $\Gamma$ to the CBM at $\Gamma$ from the data below $x = 0.3$ and dashed brown line is the fit of the data with a bowing parameter to the indirect bandgap from VBM at Γ to the CBM at $M$ from the data above $x = 0.3$.

$CaCd_2P_2$ has a direct bandgap with the conduction band minimum (CBM) at $\Gamma$ while $CaMg_2P_2$ has an indirect bandgap with the CBM at $M$. Both materials have their valence band maximum (VBM) at $\Gamma$. Therefore, it can be expected that there exists a direct-indirect crossover at a certain composition of their alloy. Figure 4A shows their band structures and the calculated bandgaps for different compositions between the end members. There is an apparent upward bowing relative to the minimum bandgap of each compound, which we also predict from the previous screening steps. If we consider the direct bandgap from the VBM and CBM both at Γ and the indirect bandgap from the VBM at Γ to the CBM at $M$ transitions separately, it can be seen that there is a crossover near $x = 0.3$. Indeed, by fitting a bowing parameter to the two compositional regions of the data above and below this crossover separately, we find excellent agreement with the calculated bandgaps (Figure 4B). For the direct transition we find a bowing parameter of 0.56 eV and for the indirect transition we find the bowing parameter is 0.38 eV. With these bowing parameters we can solve for the crossover composition (i.e. where $E_{g,direct} = E_{g,indirect}$) and find $x_{crossover} \approx 0.31$. In contrast, determination of the nature of the bandgap by direct examination of the band structure is difficult due to the folding of the Brillouin zone in supercell calculations. From the above analysis, the range of $Ca(Cd_{1-x}Mg_x)_2P_2$ compositions that have direct bandgaps becomes clear. The largest direct bandgap will occur at $x_{crossover}$ and be about 1.9 eV. The target bandgap for tandem top cell solar absorbers of 1.8 eV is at $x = 0.21$ and is direct.

As a separate check, we unfold the band structure from the SQS of $Ca(Cd_{0.75}Mg_{0.25})_2P_2$, which is near the target composition. The unfolding is independent of the analysis above and is another way to verify the direct bandgap nature of the alloy. Figure 5 displays the unfolded band structure, showing that the bandgap is indeed direct and has the desired magnitude. The bandgap at the Γ point was calculated with HSE and found to be 1.81 eV and this value was used for the rigid shift of the conduction band. As a comparison, Figure S2 shows the unfolded band structure of $Ca(Cd_{0.25}Mg_{0.75})_2P_2$ with an indirect bandgap.

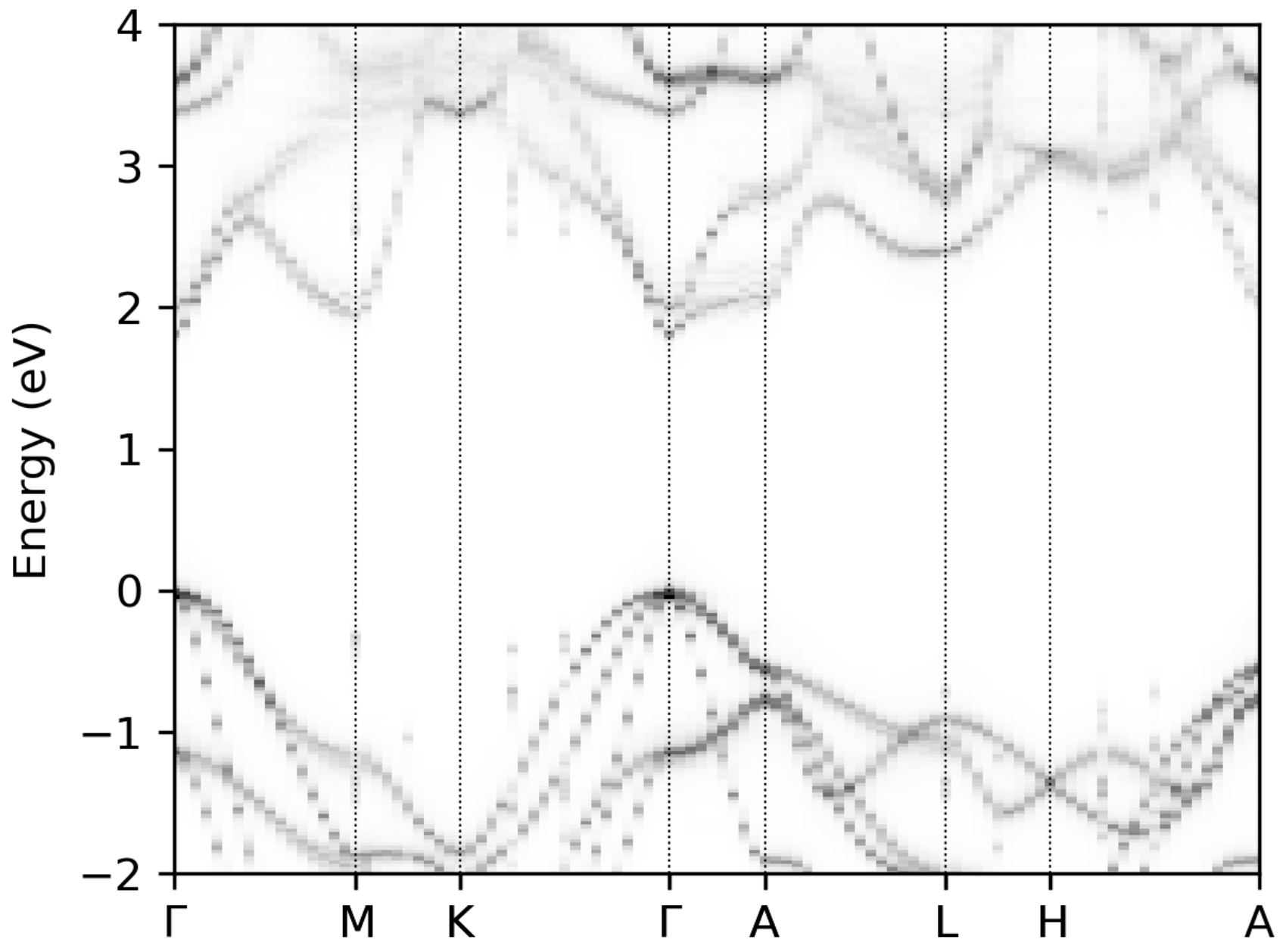

**Figure 5.** Band structure for an SQS of $Ca(Cd_{0.75}Mg_{0.25})_2P_2$. A rigid shift of the conduction band has been applied to match the bandgap at the $\Gamma$ point with HSE. The band structure unfolding and plotting were performed with Easyunfold[71].

*Synthesis of Targeted Alloys*

We have attempted the synthesis of $Ca(Cd_{1-x}Mg_x)_2P_2$ following synthetic procedures suitable to form the parent compound $CaCd_2P_2$ as a single phase. Unfortunately, we were unable to form the alloy despite favorable computed thermodynamics for its formation. Screening of a wider range of synthetic conditions may be required to achieve this goal. Alternatively, to show the synthesizability of alloys of the $AM_2P_2$, we have also attempted the synthesis of another similar alloy, $Ca(Zn_{1-x}Mg_x)_2P_2$. As stated above, we note that $Ca(Zn_{1-x}Mg_x)_2P_2$ is not an ideal candidate for tandem solar cells, but its successful growth does highlight the synthesizability of these alloys. The synthesis of $Ca(Zn_{1-x}Mg_x)_2P_2$ alloys was attempted from the flux synthesis method. Katsube et al. were also able to synthesize $AZn_2P_2$ with a Sn-flux method[72].

On the synthesized phase, we have performed SCXRD and EDS. From SCXRD, we can observe an increase in lattice parameters with the addition of Mg. Figure 6 shows EDS of the $Ca(Zn_{1-x}Mg_x)_2P_2$ sample. EDS shows a homogenous distribution of Mg in the sample. Table 2 contains the elemental composition from EDS; see Table S5 for the full composition including O, C, and Sn impurities. From this we calculate $x = \frac{x_{Mg}}{x_{Mg}+x_{Ca}} = \frac{7.7}{7.7+38.7} = 16.7\%$, though we note that this is a qualitative estimate. Note that underestimation of light elements, such as P and Mg is typical for EDS.

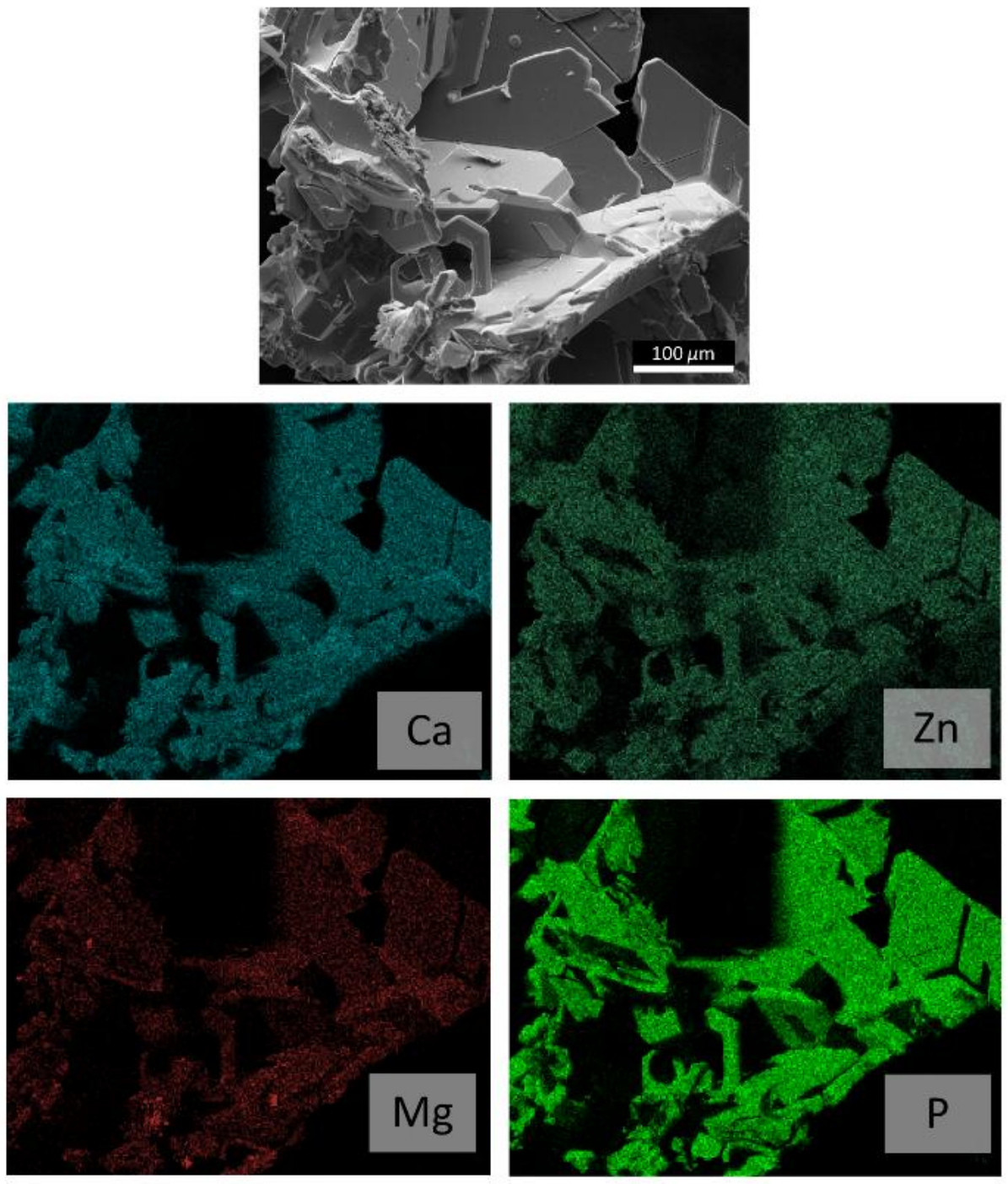

**Figure 6.** SEM image and EDS mapping of $Ca(Zn_{1-x}Mg_x)_2P_2$ crystal. The scale bar is 100 µm. Figure S3 shows the full EDS mapping including Sn, O, and C impurities.

**Table 2.** Normalized compositions of Ca, Zn, Mg, and P from EDS scan.

| Element | Atomic % |
|---|---|
| **Ca** | 20.6(6) |
| **Mg** | 7.7(7) |
| **Zn** | 39(1) |
| **P** | 33.(2) |

**Table 3.** Unit cell volume comparison.

| Method | Sample | Unit cell volume, V (Å$^3$) | ΔV (Å$^3$) |
|---|---|---|---|
| DFT (PBE) | **$CaZn_2P_2$** | 97.202 | - |
| | **$CaMg_2P_2$** | 109.021 | 11.818 |
| Experimental (PXRD) | **$CaZn_2P_2$** | 96.523 | - |
| | **$Ca(Zn_{1-x}Mg_x)_2P_2$** | 98.999 | 2.476 |

From SCXRD we have determined the lattice parameters of the $Ca(Zn_{1-x}Mg_x)_2P_2$ and find the unit cell has expanded compared to pure $CaZn_2P_2$; see Tables S5 and S6 for full results of the refinement. This is consistent with the PBE-relaxed structures of $CaZn_2P_2$ and $CaMg_2P_2$, with the latter having larger lattice parameters. For more accurate lattice parameters, a sample was ground to a powder and PXRD was performed. Figures S4 and S5 show the PXRD patterns. Table 3 shows the lattice parameters found in this investigation. Since PBE tends to overestimate lattice parameters, which makes direct comparison to experiments unrealistic, we compare them in terms of relative change. The calculations show that the unit cell volume of $CaMg_2P_2$ is 11.818 Å$^3$ larger than $CaZn_2P_2$. From PXRD, the Mg-alloyed sample has a unit cell volume 2.476 Å$^3$ larger than $CaZn_2P_2$. Assuming Vegard's Law, we estimate $x = 2.476/11.818 = 21.0\%$ and the overall composition is approximately $Ca(Zn_{0.8}Mg_{0.2})_2P_2$. This agrees well with the previous estimate from EDS. We find a similar result examining the lattice parameters $a$ and $c$ individually instead of the unit cell volume (see Table S3). Furthermore, refinement of the crystal structure from SCXRD shows that Mg and Zn jointly occupied the same crystallographic site. The refinement shows the fraction of the Mg occupying the Zn site is 32.4(8)%, 20.7(8)%, and 22.6(8)% for each of the samples. For the Zn site in the pristine $CaZn_2P_2$, the SCXRD refinement determined an average $U_{11}/U_{33}$ of 0.856, showing nearly isotropic atomic displacements. For the Mg-alloyed sample the average $U_{11}/U_{33}$ was 1.62. Tables S5 and S6 show the details of the refinement from SCXRD.

We additionally report that no degradation of the sample occurred after several weeks stored in ambient conditions. This is shown by the lack of oxide phases appearing in the PXRD patterns or in the EDS map. $CaMg_2P_2$ has never been reported, which we suspect is due to its sensitivity to air and moisture. However, the results obtained here confirm our hypothesis that a $Ca(Zn_{1-x}Mg_x)_2P_2$ alloy will be air stable with small amounts of Mg.

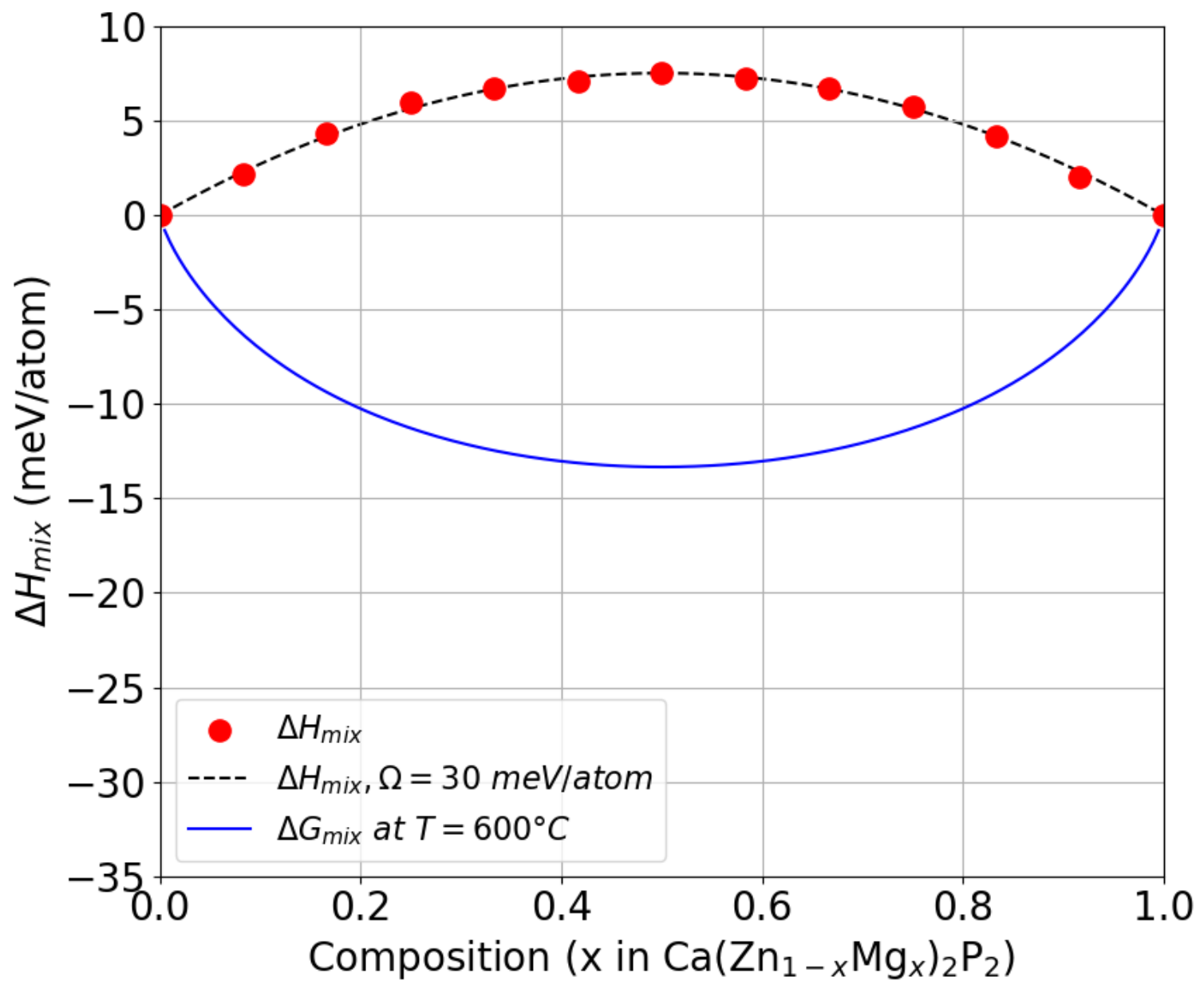

**Figure 7.** The mixing enthalpy and free energy of $Ca(Zn_{1-x}Mg_x)_2P_2$ from a composition gradient of SQSs. The red points are from the SQS calculations, the black dashed line is mixing enthalpy from the interaction parameter fit to the SQS calculations, and the blue line is the Gibbs' free energy including the effect of mixing entropy.

We have calculated the thermodynamics of $Ca(Zn_{1-x}Mg_x)_2P_2$. Figure 7 shows the calculated mixing enthalpy and the Gibb's free energy. $Ca(Zn_{1-x}Mg_x)_2P_2$ has a slightly positive interaction parameter, however its mixing enthalpy is small and can be overcome by entropic effects at a realistic synthesis temperature.

The synthesis of $Ca(Zn_{1-x}Mg_x)_2P_2$ succeeded from the Sn-flux method similar to the procedure used to synthesize $CaZn_2P_2$[72]. This is in contrast to the synthesis of $Ca(Cd_{1-x}Mg_x)_2P_2$ which was attempted from a powder synthesis method but was unsuccessful despite the similarity of the procedure to that was successful for $CaCd_2P_2$[6]. While the mixing enthalpy of $Ca(Zn_{1-x}Mg_x)_2P_2$ is slightly positive, the mixing enthalpy in $Ca(Cd_{1-x}Mg_x)_2P_2$ is actually negative as shown in Figure 3. This would further favor the formation of $Ca(Cd_{1-x}Mg_x)_2P_2$, leading us to believe that its formation is kinetically limited and perhaps the flux synthesis will produce the alloy phase. Future work will focus on refining the synthesis techniques to allow for its formation.

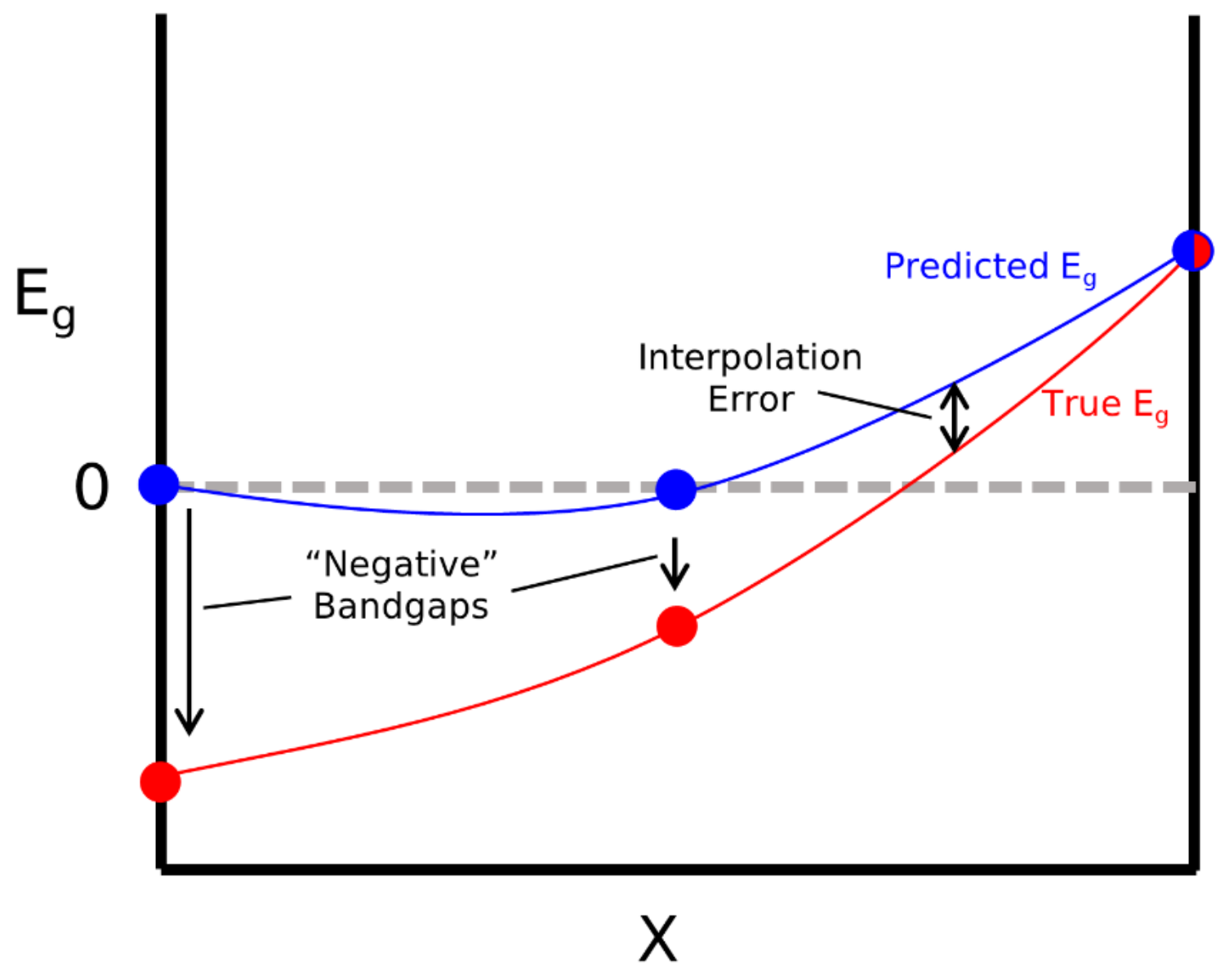

**Figure 8.** A schematic to illustrate the challenge of searching for small bandgap materials. The red curve represents the true behavior of the bandgap, whereas the blue represents the behavior as predicted by our curve fitting, when the amount of overlap of the valence and conduction bands is not accounted for in metals.

*Infrared Detector Materials*

On the other side of the absorption spectra, towards low energy, long wavelength, there is a need for materials with strong optical absorption and high carrier lifetime as well[13,73]. IR detection devices rely on p-n junctions, like solar cells, but in separate wavelengths. Among materials used in IR detectors are InSb, InAs, or mercury-cadmium telluride (MCT, $Hg_{1-x}Cd_xTe$)[23,74,75]. The far IR range (photon energies from 0.1 to 0.5 eV) is especially lacking in materials beyond MCT which has challenges in terms of growth[76]. The interesting optoelectronic properties of *$AM_2Pn_2$* could make them of interest as IR detector materials provided their bandgap can be tuned in the appropriate range.

We perform an analysis similar to the above, but screen for small bandgap materials for use as IR detectors, with a target bandgap of 0.4 eV. However, searching for small bandgap materials has an additional challenge. That is, in the bowing parameter determination, metals should ideally be treated as having a "negative" bandgap to account for the actual amount of band overlap, as shown schematically in Figure 8. If any of the compounds used to determine the bowing parameter are metals, this will lead to errors in the bandgap interpolation when solving Eq. 1 for $b$. We note that the systematic underestimation of bandgaps with PBE makes this task more challenging by making small bandgap materials seem to be metals. Despite this, we find that our method is still able to guide us towards several promising candidates. Performing the screening process again, we do find several Mg and N compounds predicted to have low bandgaps, but we believe these are due to the PBE bandgap error, so we remove them from consideration. The accuracy of the thermodynamic predictions is expected to be unaffected by this.

The only promising candidates with no miscibility gap ($\Omega < 0$) from our screening that could be interesting alloy systems for far IR detectors are $SrCd_2(Bi_xSb_{1-x})_2$ and, similarly, $CaCd_2(Bi_xSb_{1-x})_2$. $SrCd_2Sb_2$ has a fundamental bandgap of 0.34 eV and direct bandgap of 0.40 eV and $SrCd_2Bi_2$ is a metal. Therefore, it is possible to vary the bandgap of $SrCd_2(Sb_{1-x}Bi_x)_2$ from 0 to 0.34 eV. We also note that other interesting systems are $Ca(Zn_xCd_{1-x})_2Sb_2$, $SrZn_2(Sb_xAs_{1-x})_2$, and $CaZn_2(Sb_xAs_{1-x})_2$ which have small miscibility gaps and may be able to be synthesized depending on reaction conditions. It is expected that alloying $CaZn_2Sb_2$ and $CaZn_2As_2$, which respectively have bandgaps of 0.12 eV and 1.0 eV, will result in a direct bandgap material since both have direct bandgaps.

*Conclusions*

We have presented an automated computational workflow that can systematically screen for alloys, and shown its utility in finding tandem top cell solar absorbers and infrared detector materials among the class of *$AM_2Pn_2$* materials. We identify several Mg-containing *$AM_2Pn_2$* alloys, including $Ca(Zn_{1-x}Mg_x)_2N_2$, $Sr(Zn_{1-x}Mg_x)_2N_2$, $Ca(Cd_{1-x}Mg_x)_2P_2$, and $Sr(Cd_{1-x}Mg_x)_2P_2$, which are thermodynamically stable, have adequate and direct bandgaps, and require only a small amount of Mg to avoid possible air stability issues. Further SQS calculations of $Ca(Cd_{1-x}Mg_x)_2P_2$ confirm the findings from the screening and show that this alloy will have a 1.8 eV direct bandgap at $x = 0.2$. While synthesis of $Ca(Cd_{1-x}Mg_x)_2P_2$ has not yet been successful, we have shown that a $Ca(Zn_{1-x}Mg_x)_2P_2$ alloy can be synthesized with $x = 0.2$. We show that Mg is able to be incorporated into $CaZn_2P_2$ without causing air stability issues. These results are encouraging for the future realization of the targeted $Ca(Cd_{1-x}Mg_x)_2P_2$ alloy. Additionally, we have identified $SrCd_2(Bi_xSb_{1-x})_2$ and $CaCd_2(Bi_xSb_{1-x})_2$ as promising far IR detector materials from our screening. Our data can be used to design other alloys in the $AM_2Pn_2$ family and our screening method is readily applicable to design alloys in other compositional families as well.

*Acknowledgements*

This work was primarily supported by the U.S. Department of Energy, Office of Science, Basic Energy Sciences, Division of Materials Science and Engineering, Physical Behavior of Materials program under award number DE-SC0023509. All computations, syntheses, and characterizations were supported by this award unless specifically stated otherwise. This research used resources of the National Energy Research Scientific Computing Center (NERSC), a DOE Office of Science User Facility supported by the Office of Science of the U.S. Department of Energy under contract no. DE-AC02-05CH11231 using NERSC award BES ERCAP0023830. A.P. acknowledges support from a Department of Education GAANN fellowship. The analysis on infrared detector materials was supported by the United States Air Force Office of Scientific Research under award no. FA9550-22-1-0355.

*References*

(1) Yuan, Z.; Dahliah, D.; Hasan, M. R.; Kassa, G.; Pike, A.; Quadir, S.; Claes, R.; Chandler, C.; Xiong, Y.; Kyveryga, V.; others. Discovery of the Zintl-Phosphide BaCd2P2 as a Long Carrier Lifetime and Stable Solar Absorber. *Joule* **2024**, *8* (5), 1412–1429.

(2) Kassa, G.; Yuan, Z.; Hasan, M. R.; Esparza, G. L.; Fenning, D. P.; Hautier, G.; Kovnir, K.; Liu, J. BaCd2P2: A Defect-Resistant "GaAs." **2025**, 1–35.

(3) Hautzinger, M. P.; Quadir, S.; Feingold, B.; Seban, R.; Thornton, A. J.; Dutta, N. S.; Norman, A. G.; Leahy, I. A.; Hasan, M. R.; Kovnir, K. A.; Reid, O. G.; Larson, B. W.; Luther, J. M.; Beard, M. C.; Bauers, S. R. Synthesis and Characterization of Zintl-Phase BaCd 2 P 2 Quantum Dots for Optoelectronic Applications. *ACS Nano* **2025**, *19* (12), 12345–12353. https://doi.org/10.1021/acsnano.5c02271.

(4) Quadir, S.; Yuan, Z.; Esparza, G. L.; Dugu, S.; Mangum, J. S.; Pike, A.; Hasan, M. R.; Kassa, G.; Wang, X.; Coban, Y.; Liu, J.; Kovnir, K.; Fenning, D. P.; Reid, O. G.; Zakutayev, A.; Hautier, G.; Bauers, S. R. Low-Temperature Synthesis of Stable CaZn2P2 Zintl Phosphide Thin Films as Candidate Top Absorbers. *Adv Energy Mater* **2024**, *14* (44), 2402640. https://doi.org/10.1002/aenm.202402640.

(5) Esparza, G. L.; Yuan, Z.; Hasan, M. R.; Coban, Y.; Kassa, G.; Devalla, V. S.; Nivarty, T.; Palmer, J. R.; Liu, J.; Kovnir, K.; Hautier, G.; Fenning, D. P. CaCd2P2: A Visible-Light Absorbing Zintl Phosphide Stable under Photoelectrochemical Water Oxidation. **2025**.

(6) Pike, A.; Yuan, Z.; Kassa, G.; Hasan, M. R.; Goswami, S.; Dugu, S.; Quadir, S.; Zakutayev, A.; Bauers, S. R.; Kovnir, K.; Liu, J.; Hautier, G. Map of the Zintl AM 2 Pn 2 Compounds: Influence of Chemistry on Stability and Electronic Structure. *Chemistry of Materials* **2025**, *37* (13), 4684–4694. https://doi.org/10.1021/acs.chemmater.5c00353.

(7) NREL. *Best Research-Cell Efficiency Chart*. https://www.nrel.gov/pv/cell-efficiency (accessed 2025-01-06).

(8) Yu, M. L.; Los, A.; Xiong, G. Thin Film Absorbers for Tandem Solar Cells: An Industrial Perspective. *Journal of Physics: Energy* **2023**, *5* (4), 042002. https://doi.org/10.1088/2515-7655/acff18.

(9) Alberi, K.; Berry, J. J.; Cordell, J. J.; Friedman, D. J.; Geisz, J. F.; Kirmani, A. R.; Larson, B. W.; McMahon, W. E.; Mansfield, L. M.; Ndione, P. F.; Owen-Bellini, M.; Palmstrom, A. F.; Reese, M. O.; Reese, S. B.; Steiner, M. A.; Tamboli, A. C.; Theingi, S.; Warren, E. L. A Roadmap for Tandem Photovoltaics. *Joule* **2024**, *8* (3), 658–692. https://doi.org/10.1016/j.joule.2024.01.017.

(10) Aydin, E.; Allen, T. G.; De Bastiani, M.; Razzaq, A.; Xu, L.; Ugur, E.; Liu, J.; De Wolf, S. Pathways toward Commercial Perovskite/Silicon Tandem Photovoltaics. *Science (1979)* **2024**, *383* (6679), 1–13. https://doi.org/10.1126/science.adh3849.

(11) Peplow, M. A New Kind of Solar Cell Is Coming: Is It the Future of Green Energy? *Nature* **2023**, *623* (7989), 902–905. https://doi.org/10.1038/d41586-023-03714-y.

(12) Hull, M.; Rousset, J.; Nguyen, V. S.; Grand, P.-P.; Oberbeck, L. Prospective Techno-Economic Analysis of 4T and 2T Perovskite on Silicon Tandem Photovoltaic Modules at GW-Scale Production. *Solar RRL* **2023**, *7* (23), 1–7. https://doi.org/10.1002/solr.202300503.

(13) Kinch, M. A. Fundamental Physics of Infrared Detector Materials. *J Electron Mater* **2000**, *29* (6), 809–817. https://doi.org/10.1007/s11664-000-0229-7.

(14) Ablekim, T.; Duenow, J. N.; Perkins, C. L.; Moseley, J.; Zheng, X.; Bidaud, T.; Frouin, B.; Collin, S.; Reese, M. O.; Amarasinghe, M.; Colegrove, E.; Johnston, S.; Metzger, W. K. Exceeding 200 Ns Lifetimes in Polycrystalline CdTe Solar Cells. *Solar RRL* **2021**, *5* (8), 1–7. https://doi.org/10.1002/solr.202100173.

(15) Onno, A.; Reich, C.; Li, S.; Danielson, A.; Weigand, W.; Bothwell, A.; Grover, S.; Bailey, J.; Xiong, G.; Kuciauskas, D.; Sampath, W.; Holman, Z. C. Understanding What Limits the Voltage of Polycrystalline CdSeTe Solar Cells. *Nat Energy* **2022**, *7* (5), 400–408. https://doi.org/10.1038/s41560-022-00985-z.

(16) Jackson, P.; Hariskos, D.; Lotter, E.; Paetel, S.; Wuerz, R.; Menner, R.; Wischmann, W.; Powalla, M. New World Record Efficiency for Cu(In,Ga)Se 2 Thin-film Solar Cells beyond 20%. *Progress in Photovoltaics: Research and Applications* **2011**, *19* (7), 894–897. https://doi.org/10.1002/pip.1078.

(17) Quadir, S.; Qorbani, M.; Lai, Y.-R.; Sabbah, A.; Thong, H.; Hayashi, M.; Chen, C.; Chen, K.; Chen, L. Impact of Cation Substitution in (Ag x Cu 1− x ) 2 ZnSnSe 4 Absorber-Based Solar Cells toward 10% Efficiency: Experimental and Theoretical Analyses. *Solar RRL* **2021**, *5* (10), 1–9. https://doi.org/10.1002/solr.202100441.

(18) Keller, J.; Kiselman, K.; Donzel-Gargand, O.; Martin, N. M.; Babucci, M.; Lundberg, O.; Wallin, E.; Stolt, L.; Edoff, M. High-Concentration Silver Alloying and Steep Back-Contact Gallium Grading Enabling Copper Indium Gallium Selenide Solar Cell with 23.6% Efficiency. *Nat Energy* **2024**, *9* (4), 467–478. https://doi.org/10.1038/s41560-024-01472-3.

(19) Gershon, T.; Lee, Y. S.; Antunez, P.; Mankad, R.; Singh, S.; Bishop, D.; Gunawan, O.; Hopstaken, M.; Haight, R. Photovoltaic Materials and Devices Based on the Alloyed Kesterite Absorber (Ag x Cu 1– x ) 2 ZnSnSe 4. *Adv Energy Mater* **2016**, *6* (10), 4–10. https://doi.org/10.1002/aenm.201502468.

(20) Qi, Y.; Tian, Q.; Meng, Y.; Kou, D.; Zhou, Z.; Zhou, W.; Wu, S. Elemental Precursor Solution Processed (Cu 1– x Ag x ) 2 ZnSn(S,Se) 4 Photovoltaic Devices with over 10% Efficiency. *ACS Appl Mater Interfaces* **2017**, *9* (25), 21243–21250. https://doi.org/10.1021/acsami.7b03944.

(21) Agarwal, S.; Vincent, K. C.; Agrawal, R. From Synthesis to Application: A Review of BaZrS 3 Chalcogenide Perovskites. *Nanoscale* **2025**, *17* (8), 4250–4300. https://doi.org/10.1039/D4NR03880K.

(22) Meng, W.; Saparov, B.; Hong, F.; Wang, J.; Mitzi, D. B.; Yan, Y. Alloying and Defect Control within Chalcogenide Perovskites for Optimized Photovoltaic Application. *Chemistry of Materials* **2016**, *28* (3), 821–829. https://doi.org/10.1021/acs.chemmater.5b04213.

(23) Lei, W.; Antoszewski, J.; Faraone, L. Progress, Challenges, and Opportunities for HgCdTe Infrared Materials and Detectors. *Appl Phys Rev* **2015**, *2* (4), 041303. https://doi.org/10.1063/1.4936577.

(24) Lawson, W. D.; Putley, E. H.; Nielsen, S.; Young, A. S. Preparation and Properties of HgTe and Mixed Crystals of HgTe-CdTe. *Journal of Physical Chemistry of Solids* **1959**, *9*, 325–329.

(25) Imasato, K.; Kang, S. D.; Ohno, S.; Snyder, G. J. Band Engineering in Mg3Sb2 by Alloying with Mg3Bi2 for Enhanced Thermoelectric Performance. *Mater Horiz* **2018**, *5* (1), 59–64. https://doi.org/10.1039/c7mh00865a.

(26) Imasato, K.; Wood, M.; Anand, S.; Kuo, J. J.; Snyder, G. J. Understanding the High Thermoelectric Performance of Mg3Sb2-Mg3Bi2 Alloys. *Advanced Energy and Sustainability Research* **2022**, *3* (6). https://doi.org/10.1002/aesr.202100208.

(27) Ponnambalam, V.; Morelli, D. T. On the Thermoelectric Properties of Zintl Compounds Mg3Bi2−x Pn x (Pn = P and Sb). *J Electron Mater* **2013**, *42* (7), 1307–1312. https://doi.org/10.1007/s11664-012-2417-7.

(28) Imasato, K.; Anand, S.; Gurunathan, R.; Snyder, G. J. The Effect of Mg3As2 Alloying on the Thermoelectric Properties of N-Type Mg3(Sb, Bi)2. *Dalton Transactions* **2021**, *50* (27), 9376–9382. https://doi.org/10.1039/d1dt01600h.

(29) Ahmadpour, F.; Kolodiazhnyi, T.; Mozharivskyj, Y. Structural and Physical Properties of Mg3−xZnxSb2 (X=0–1.34). *J Solid State Chem* **2007**, *180* (9), 2420–2428. https://doi.org/10.1016/j.jssc.2007.06.011.

(30) Tortorella, D. S.; Ghosh, K.; Bobev, S. Realization of a Trigonal Mg3–Zn P2 Intermediate Solid Solution between the Binary Cubic Mg3P2 and Tetragonal Zn3P2 End Members. *J Solid State Chem* **2025**, *344* (September 2024), 125184. https://doi.org/10.1016/j.jssc.2025.125184.

(31) Katsube, R.; Nose, Y. Experimental Investigation of Phase Equilibria around a Ternary Compound Semiconductor Mg(Mg Zn1-)2P2 in the Mg–P–Zn System at 300 °C Using Sn Flux. *J Solid State Chem* **2019**, *280* (April), 120983. https://doi.org/10.1016/j.jssc.2019.120983.

(32) Wood, M.; Aydemir, U.; Ohno, S.; Snyder, G. J. Observation of Valence Band Crossing: The Thermoelectric Properties of CaZn2Sb2 –CaMg2Sb2 Solid Solution. *J Mater Chem A Mater* **2018**, *6* (20), 9437–9444. https://doi.org/10.1039/C8TA02250J.

(33) Wu, L.; Zhou, Z.; Han, G.; Zhang, B.; Yu, J.; Wang, H.; Chen, Y.; Lu, X.; Wang, G.; Zhou, X. Realizing High Thermoelectric Performance in P-Type CaZn2Sb2-Alloyed Mg3Sb2-Based Materials via Band and Point Defect Engineering. *Chemical Engineering Journal* **2023**, *475* (September), 145988. https://doi.org/10.1016/j.cej.2023.145988.

(34) Wang, J.; Mark, J.; Woo, K. E.; Voyles, J.; Kovnir, K. Chemical Flexibility of Mg in Pnictide Materials: Structure and Properties Diversity. *Chemistry of Materials* **2019**, *31* (20), 8286–8300. https://doi.org/10.1021/acs.chemmater.9b03740.

(35) Tsuji, M.; Hiramatsu, H.; Hosono, H. Tunable Light Emission through the Range 1.8–3.2 EV and p-Type Conductivity at Room Temperature for Nitride Semiconductors, Ca(Mg1–x Znx)2N2 (X= 0–1). *Inorg Chem* **2019**, *58* (18), 12311–12316. https://doi.org/10.1021/acs.inorgchem.9b01811.

(36) Jeong, J.; Shim, D.; Choi, M.-H.; Yunxiu, Z.; Kim, D.-H.; Ok, K. M.; You, T.-S. Golden Ratio of the R+/r- for the Structure-Selectivity in the Thermoelectric BaZn2-XCdxSb2 System. *J Alloys Compd* **2024**, *1002* (June), 175272. https://doi.org/10.1016/j.jallcom.2024.175272.

(37) Yang, J.; Manganaris, P.; Mannodi-Kanakkithodi, A. A High-Throughput Computational Dataset of Halide Perovskite Alloys. *Digital Discovery* **2023**, *2* (3), 856–870. https://doi.org/10.1039/D3DD00015J.

(38) Woods-Robinson, R.; Horton, M. K.; Persson, K. A. A Method to Computationally Screen for Tunable Properties of Crystalline Alloys. **2022**. https://doi.org/10.1016/j.patter.2023.100723.

(39) Bhattacharya, S.; Madsen, G. K. H. High-Throughput Exploration of Alloying as Design Strategy for Thermoelectrics. *Phys Rev B* **2015**, *92* (8), 085205. https://doi.org/10.1103/PhysRevB.92.085205.

(40) Kresse, G.; Furthmüller, J. Efficient Iterative Schemes for Ab Initio Total-Energy Calculations Using a Plane-Wave Basis Set. *Phys. Rev. B* **1996**, *54* (16), 11169–11186. https://doi.org/10.1103/PhysRevB.54.11169.

(41) Kresse, G.; Joubert, D. From Ultrasoft Pseudopotentials to the Projector Augmented-Wave Method. *Phys. Rev. B* **1999**, *59* (3), 1758–1775. https://doi.org/10.1103/PhysRevB.59.1758.

(42) Kresse, G.; Furthmüller, J. Efficiency of Ab-Initio Total Energy Calculations for Metals and Semiconductors Using a Plane-Wave Basis Set. *Comput Mater Sci* **1996**, *6* (1), 15–50. https://doi.org/10.1016/0927-0256(96)00008-0.

(43) Kresse, G.; Hafner, J. Ab Initio Molecular-Dynamics Simulation of the Liquid-Metal--Amorphous-Semiconductor Transition in Germanium. *Phys. Rev. B* **1994**, *49* (20), 14251–14269. https://doi.org/10.1103/PhysRevB.49.14251.

(44) Mathew, K.; Montoya, J. H.; Faghaninia, A.; Dwarakanath, S.; Aykol, M.; Tang, H.; Chu, I.; Smidt, T.; Bocklund, B.; Horton, M.; Dagdelen, J.; Wood, B.; Liu, Z.-K.; Neaton, J.; Ong, S. P.; Persson, K.; Jain, A. Atomate: A High-Level Interface to Generate, Execute, and Analyze Computational Materials Science Workflows. *Comput Mater Sci* **2017**, *139*, 140–152. https://doi.org/https://doi.org/10.1016/j.commatsci.2017.07.030.

(45) Perdew, J. P.; Burke, K.; Ernzerhof, M. Generalized Gradient Approximation Made Simple. *Phys. Rev. Lett.* **1996**, *77* (18), 3865–3868. https://doi.org/10.1103/PhysRevLett.77.3865.

(46) Setyawan, W.; Curtarolo, S. High-Throughput Electronic Band Structure Calculations: Challenges and Tools. *Comput Mater Sci* **2010**, *49* (2), 299–312. https://doi.org/10.1016/j.commatsci.2010.05.010.

(47) Ong, S. P.; Richards, W. D.; Jain, A.; Hautier, G.; Kocher, M.; Cholia, S.; Gunter, D.; Chevrier, V. L.; Persson, K. A.; Ceder, G. Python Materials Genomics (Pymatgen): A Robust, Open-Source Python Library for Materials Analysis. *Comput Mater Sci* **2013**, *68*, 314–319. https://doi.org/10.1016/j.commatsci.2012.10.028.

(48) Furness, J. W.; Kaplan, A. D.; Ning, J.; Perdew, J. P.; Sun, J. Accurate and Numerically Efficient R2SCAN Meta-Generalized Gradient Approximation. *J Phys Chem Lett* **2020**, *11* (19), 8208–8215. https://doi.org/10.1021/acs.jpclett.0c02405.

(49) Zhang, Y.; Ramasamy, A.; Pokharel, K.; Kothakonda, M.; Xiao, B.; Furness, J. W.; Ning, J.; Zhang, R.; Sun, J. Advances and Challenges of SCAN and R2SCAN Density Functionals in Transition-Metal Compounds. *WIREs Computational Molecular Science* **2025**, *15* (2). https://doi.org/10.1002/wcms.70007.

(50) Kothakonda, M.; Kaplan, A. D.; Isaacs, E. B.; Bartel, C. J.; Furness, J. W.; Ning, J.; Wolverton, C.; Perdew, J. P.; Sun, J. Testing the R2SCAN Density Functional for the Thermodynamic Stability of Solids with and without a van Der Waals Correction. *ACS Materials Au* **2023**, *3* (2), 102–111. https://doi.org/10.1021/acsmaterialsau.2c00059.

(51) Kingsbury, R.; Gupta, A. S.; Bartel, C. J.; Munro, J. M.; Dwaraknath, S.; Horton, M.; Persson, K. A. Performance Comparison of R2SCAN and SCAN MetaGGA Density Functionals for Solid Materials via an Automated, High-Throughput Computational Workflow. *Phys. Rev. Mater.* **2022**, *6* (1), 13801. https://doi.org/10.1103/PhysRevMaterials.6.013801.

(52) Krukau, A. V.; Vydrov, O. A.; Izmaylov, A. F.; Scuseria, G. E. Influence of the Exchange Screening Parameter on the Performance of Screened Hybrid Functionals. *Journal of Chemical Physics* **2006**, *125* (22). https://doi.org/10.1063/1.2404663.

(53) Wei, S.-H.; Zhang, S. B.; Zunger, A. First-Principles Calculation of Band Offsets, Optical Bowings, and Defects in CdS, CdSe, CdTe, and Their Alloys. *J Appl Phys* **2000**, *87* (3), 1304–1311. https://doi.org/10.1063/1.372014.

(54) Moon, C.-Y.; Wei, S.-H.; Zhu, Y. Z.; Chen, G. D. Band-Gap Bowing Coefficients in Large Size-Mismatched II-VI Alloys: First-Principles Calculations. *Phys Rev B* **2006**, *74* (23), 233202. https://doi.org/10.1103/PhysRevB.74.233202.

(55) Chen, S.; Gong, X. G.; Wei, S.-H. Band-Structure Anomalies of the Chalcopyrite Semiconductors CuGaX2 versus AgGaX2 (X=S and Se) and Their Alloys. *Phys Rev B* **2007**, *75* (20), 205209. https://doi.org/10.1103/PhysRevB.75.205209.

(56) Jain, A.; Ong, S. P.; Hautier, G.; Chen, W.; Richards, W. D.; Dacek, S.; Cholia, S.; Gunter, D.; Skinner, D.; Ceder, G.; Persson, K. A. Commentary: The Materials Project: A Materials Genome Approach to Accelerating Materials Innovation. *APL Mater* **2013**, *1* (1), 11002. https://doi.org/10.1063/1.4812323.

(57) Horton, M. K.; Huck, P.; Yang, R. X.; Munro, J. M.; Dwaraknath, S.; Ganose, A. M.; Kingsbury, R. S.; Wen, M.; Shen, J. X.; Mathis, T. S.; Kaplan, A. D.; Berket, K.; Riebesell, J.; George, J.; Rosen, A. S.; Spotte-Smith, E. W. C.; McDermott, M. J.; Cohen, O. A.; Dunn, A.; Kuner, M. C.; Rignanese, G.; Petretto, G.; Waroquiers, D.; Griffin, S. M.; Neaton, J. B.; Chrzan, D. C.; Asta, M.; Hautier, G.; Cholia, S.; Ceder, G.; Ong, S. P.; Jain, A.; Persson, K. A. Accelerated Data-Driven Materials Science with the Materials Project. *Nat Mater* **2025**, No. Ml. https://doi.org/10.1038/s41563-025-02272-0.

(58) Zunger, A.; Wei, S.-H.; Ferreira, L. G.; Bernard, J. E. Special Quasirandom Structures. *Phys Rev Lett* **1990**, *65* (3), 353–356.

(59) Wei, S.-H.; Ferreira, L. G.; Bernard, J. E.; Zunger, A. Electronic Properties of Random Alloys: Special Quasirandom Structures. *Phys Rev B* **1990**, *42* (15), 9622–9649. https://doi.org/10.1103/PhysRevB.42.9622.

(60) Walle, A. Van De; Tiwary, P.; Jong, M. De; Olmsted, D. L.; Asta, M.; Dick, A.; Shin, D.; Wang, Y.; Chen, L.; Liu, Z.; Carlo, M. Efficient Stochastic Generation of Special Quasirandom Structures. *Calphad* **2013**, *42*, 13–18.

(61) van de Walle, A.; Asta, M.; Ceder, G. The Alloy Theoretic Automated Toolkit: A User Guide. *Calphad* **2002**, *26* (4), 539–553. https://doi.org/10.1016/S0364-5916(02)80006-2.

(62) Wei, S.-H.; Zunger, A. Giant and Composition-Dependent Optical Bowing Coefficient in GaAsN Alloys. *Phys Rev Lett* **1996**, *76* (4), 664–667. https://doi.org/10.1103/PhysRevLett.76.664.

(63) Wei, S.-H.; Zunger, A. Band Offsets and Optical Bowings of Chalcopyrites and Zn-Based II-VI Alloys. *J Appl Phys* **1995**, *78* (6), 3846–3856. https://doi.org/10.1063/1.359901.

(64) CrysalisPro. Rigaku Corporation: Oxford, United Kingdon 2025.

(65) Sheldrick, G. M. SHELXT – Integrated Space-Group and Crystal-Structure Determination. *Acta Crystallogr A Found Adv* **2015**, *71* (1), 3–8. https://doi.org/10.1107/S2053273314026370.

(66) Sheldrick, G. M. SHELXL. University of Götttingen: Göttingen, Germany 2014.

(67) Dolomanov, O. V.; Bourhis, L. J.; Gildea, R. J.; Howard, J. A. K.; Puschmann, H. OLEX2 : A Complete Structure Solution, Refinement and Analysis Program. *J Appl Crystallogr* **2009**, *42* (2), 339–341. https://doi.org/10.1107/S0021889808042726.

(68) Deller, K.; Eisenmann, B. Ternäre Erdalkali-Element(V)-Verbindungen AMg 2 B 2 Mit A = Ca, Sr, Ba Und B = As, Sb, Bi / Ternary Alkaline Earth-Element(V)-Compounds AMg 2 B 2 with A = Ca, Sr, Ba and B = As, Sb, Bi . *Zeitschrift für Naturforschung B* **1977**, *32* (6), 612–616. https://doi.org/10.1515/znb-1977-0602.

(69) Hinuma, Y.; Hatakeyama, T.; Kumagai, Y.; Burton, L. A.; Sato, H.; Muraba, Y.; Iimura, S.; Hiramatsu, H.; Tanaka, I.; Hosono, H.; Oba, F. Discovery of Earth-Abundant Nitride Semiconductors by Computational Screening and High-Pressure Synthesis. *Nat Commun* **2016**, *7* (1), 11962. https://doi.org/10.1038/ncomms11962.

(70) Tsuji, M.; Hanzawa, K.; Kinjo, H.; Hiramatsu, H.; Hosono, H. Heteroepitaxial Thin-Film Growth of a Ternary Nitride Semiconductor CaZn2N2. *ACS Appl Electron Mater* **2019**, *1* (8), 1433–1438. https://doi.org/10.1021/acsaelm.9b00248.

(71) Zhu, B.; Kavanagh, S. R.; Scanlon, D. Easyunfold: A Python Package for Unfolding Electronic Band Structures. *J Open Source Softw* **2024**, *9* (93), 5974. https://doi.org/10.21105/joss.05974.

(72) Katsube, R.; Nose, Y. Synthesis of Alkaline-Earth Zintl Phosphides M Zn 2 P 2 ( M = Ca, Sr, Ba) from Sn Solutions. *High Temperature Materials and Processes* **2022**, *41* (1), 8–15. https://doi.org/10.1515/htmp-2022-0019.

(73) Rolgalski, A.; Kopytko, M.; Martyniuk, P. *Antimonide-Based Infrared Detectors: A New Perspective*; SPIE: Bellingham, Wahington USA, 2018.

(74) Rogalski, A. HgCdTe Infrared Detector Material: History, Status and Outlook. *Reports on Progress in Physics* **2005**, *68* (10), 2267–2336. https://doi.org/10.1088/0034-4885/68/10/R01.

(75) Rogalski, A.; Martyniuk, P.; Kopytko, M. InAs/GaSb Type-II Superlattice Infrared Detectors: Future Prospect. *Appl Phys Rev* **2017**, *4* (3). https://doi.org/10.1063/1.4999077.

(76) Vella, J. H.; Huang, L.; Eedugurala, N.; Mayer, K. S.; Ng, T. N.; Azoulay, J. D. Broadband Infrared Photodetection Using a Narrow Bandgap Conjugated Polymer. *Sci Adv* **2021**, *7* (24), 6–11. https://doi.org/10.1126/sciadv.abg2418.

# Supplemental Information for

# Alloying to Tune the Bandgap of the *AM₂Pn₂* Zintl Compounds

Andrew Pike[1], Zhenkun Yuan[1], Muhammad Rubaiat Hasan[2], Smitakshi Goswami[1,3], Krishanu Samanta[4], Miguel I. Gonzalez[4], Jifeng Liu[1], Kirill Kovnir[2,5], Geoffroy Hautier[1,6,7]

1 Thayer School of Engineering, Dartmouth College, Hanover NH, 03755, USA

2 Department of Chemistry, Iowa State University, Ames, Iowa 50011, USA

3 Department of Physics, Dartmouth College, Hanover NH, 03755, USA

4 Department of Chemistry, Dartmouth College, Hanover NH, 03755, USA

5 Ames National Laboratory, U.S. Department of Energy, Ames, Iowa 50011, USA

6 Department of Materials Science and NanoEngineering, Rice University, Houston, TX 77005, USA

7 Rice Institute of Advanced Materials, Rice University, Houston, TX 77005, USA

Corresponding author email: geoffroy.hautier@rice.edu

*Literature Comparison*

Here, we summarize the other reports of $AM_2Pn_2$ alloys. One of the most notable alloys is the $Mg_3Sb_{1-x}Bi_x$[1–3] which has attracted attention for its thermoelectric properties and has been synthesized from direct reaction of the elemental powders and with ball milling. Additionally, a pseudoternary alloy with the addition of As has been reported, namely $Mg_3((Sb_{0.5}Bi_{0.5})_{1-x}As_x)_2$[4] from ball milling. $Mg_3Bi_{1-x}P_x$[2] has also been synthesized, another alloy of $Mg_3Pn_2$ with mixing on the *Pn* site. Other $Mg_3Pn_2$ alloys have incorporated Zn into the *M* site: $Mg(Mg_xZn_{1-x})_2P_2$[5,6], $Mg_{3-x}Zn_xSb_2$[7], and $(Mg_{1-x}Ca_x)(Mg_{1-x}Zn_x)_2Sb_2$[8]. There are also two antimonides with separate *A* and *M* cations: $Ba(Zn_{1-x}Cd_x)_2Sb_2$[9] from a Pb-flux synthesis and $Ca(Zn_{1-x}Mg_x)_2Sb_2$[10] from ball milling.

Particularly relevant to this investigation is the report of the synthesis of $Ca(Mg_{1-x}Zn_x)_2N_2$ and characterization of its bandgap by Tsuji et al[11]. The alloy was able to be formed through the entire range of x, although $x > 0.12$ required synthesis in a high pressure apparatus. This matches with our prediction of an $E_{hull} < 0$. Their results on the bandgap of this system are an ideal benchmark for our bowing parameter estimates. For this system we have calculated $b = 0.31\ eV$. The measurements in Tsuji et al., show no obvious bowing. Our calculated b is relatively small, for example the maximum $\Delta Eg$ would be less than 0.08 eV at $x = 0.5$ which is within the error of the measurements of Tsuji et al., so the measurements are in agreement. It is also noted that for $x < 0.5$ (high Mg concentration), the samples become unstable in air.

Table S1 summarizes the experimental results for $AM_2Pn_2$ alloys. Overall, we find good agreement for the $E_{hull}$ and the synthesized phases in that they are predicted to be close to the hull. The exception is for $Mg_3(Bi_{1-x}P_x)_2$ which has an $E_{hull}$ of 67 meV/atom. We also note that ball milling is a high energy process that may be used to form metastable phases.

Table S1 – Reported $AM_2Pn_2$ alloys and calculated $E_{hull}$

| Alloy family | Compositions | Synthesis method | $E_{hull}$ (meV/atom) (x=0.5) | Source |
|---|---|---|---|---|
| **$Mg_3(Sb_{1-x}Bi_x)_2$** | x=0, 0.25, 0.5 | ball milling | 14 | [12] |
| **$Mg_3(Sb_{1-x}Bi_x)_2$** | up to x=1 | solid state | 14 | [2] |
| **$Mg(Mg_xZn_{1-x})_2P_2$** | x=0 to 0.55 | Sn-flux | 16 | [5] |
| **$Mg(Mg_xZn_{1-x})_2P_2$** | X=0.45 to 1 | Sn-flux | 16 | [6] |
| **$Mg_3((Sb_{0.5}Bi_{0.5})_{1-x}As_x)_2$** | up to x=0.15 | ball milling | 45(for $Mg_3SbAs$)<br>49(for $Mg_3BiAs$) | [4] |
| **$Mg_3(Bi_{1-x}P_x)_2$** | up to x=0.5 | solid state | 67 | [2] |
| **$Mg_{3-x}Zn_xSb_2$** | x=0–1.34 | solid state | 4 | [7] |
| **$(Mg_{1-x}Ca_x)(Mg_{1-x}Zn_x)_2Sb_2$** | x = 0, 0.1, 0.2, 0.3, 0.4, 0.5 | ball milling | <1 (for $CaZnMgSb_2$) | [8] |
| **$Ca(Zn_{1-x}Mg_x)_2Sb_2$** | x=0 to 1 | ball milling | <1 | [10] |
| **$Ca(Zn_{1-x}Mg_x)_2N_2$** | x=0 to 1 | Solid state powder synthesis | -14 | [11] |
| **$Ba(Zn_{1-x}Cd_x)_2Sb_2$** | x=.92 and 1 | Pb-flux | 26 | [9] |

*Error cancellation in bowing parameter*:

$$E_g(x) = xE_{g,A} + (1-x)E_{g,A} - bx(1-x)$$

At x=0.5:

$$b = 4\left(\frac{(E_{g,A} + E_{g,B})}{2} - E_{g,x=0.5}\right)$$

Show error cancellation by considering:

$$E_{g,HSE} = E_{g,PBE} + \delta$$

And

$$b_{HSE} = 4\left(\frac{(E_{g,A,HSE} + E_{g,B,HSE})}{2} - E_{g,x=0.5,HSE}\right)$$

And

$$b_{PBE} = 4\left(\frac{(E_{g,A,PBE} + E_{g,B,PBE})}{2} - E_{g,x=0.5,PBE}\right)$$

So

$$b_{HSE} = 4\left(\frac{\left(E_{g,A,PBE} + \delta_A + E_{g,B,PBE} + \delta_B\right)}{2} - E_{g,x=0.5,PBE} + \delta_{x=0.5}\right)$$
$$= 4\left(\frac{\left(E_{g,A,PBE} + E_{g,B,PBE}\right)}{2} - E_{g,x=0.5,PBE}\right) + 4\left(\frac{(\delta_A + \delta_B)}{2} - \delta_{x=0.5}\right)$$
$$= b_{PBE} + 4\left(\frac{(\delta_A + \delta_B)}{2} - \delta_{x=0.5}\right)$$

If $\delta_A \approx \delta_B \approx \delta_{x=0.5}$, then

$$b_{HSE} \approx b_{PBE}$$

Showing that the bowing parameter calculated from PBE and is an acceptable approximation when applied to HSE bandgaps, avoiding calculation of the (large) SQS in HSE so only the (small unit cell) endmembers need to be calculated with HSE.

Table S2 – Comparison of experimental and calculated bowing parameters for III-V semiconductor alloys.

| Alloy | Transition (VBM to CBM) | Experimental b (eV) | Calculated b (eV) |
|---|---|---|---|
| **$Al_{1-x}Ga_xP$** | $\Gamma \rightarrow X$ | 0.13 | 0.15 |
| **$Al_{1-x}In_xP$** | $\Gamma \rightarrow X$ | 0.38 | 0.72 |
| **$In_{1-x}Ga_xP$** | $\Gamma \rightarrow \Gamma$ | 0.65 | 0.78 |

As a benchmark of the bowing parameter determination used in the high throughput screening, the same method was used for alloys with known bowing parameters to compare to. 80 atom SQS were used for the zincblende structure. Alloys with large bandgaps were selected to avoid issues with zero bandgaps as illustrated by Figure 8. Due to band folding in supercell calculations such as SQS, the nature of the bandgap is not straightforward, so the minimum bandgap for $Al_{1-x}In_xP$ and $In_{1-x}Ga_xP$ was assigned to a k-point in the CBM according to the results in Nicklas and Wilkins[14]. $Al_{1-x}Ga_xP$'s CBM at x=0.5 was assumed to be at $X$ according to Vurgaftman et al.[13] Overall the computed bowing parameters match well with the literature values. While the computed bowing parameters for $Al_{1-x}Ga_xP$ and $In_{1-x}Ga_xP$ agree quite well with their respective experimental values, $Al_{1-x}In_xP$ has the largest difference. However, this difference is still tolerable and would at most lead to a discrepancy of about 85 meV for the bandgap at x=0.5 (where the error would be the largest). For the high throughput screening this level of accuracy is acceptable.

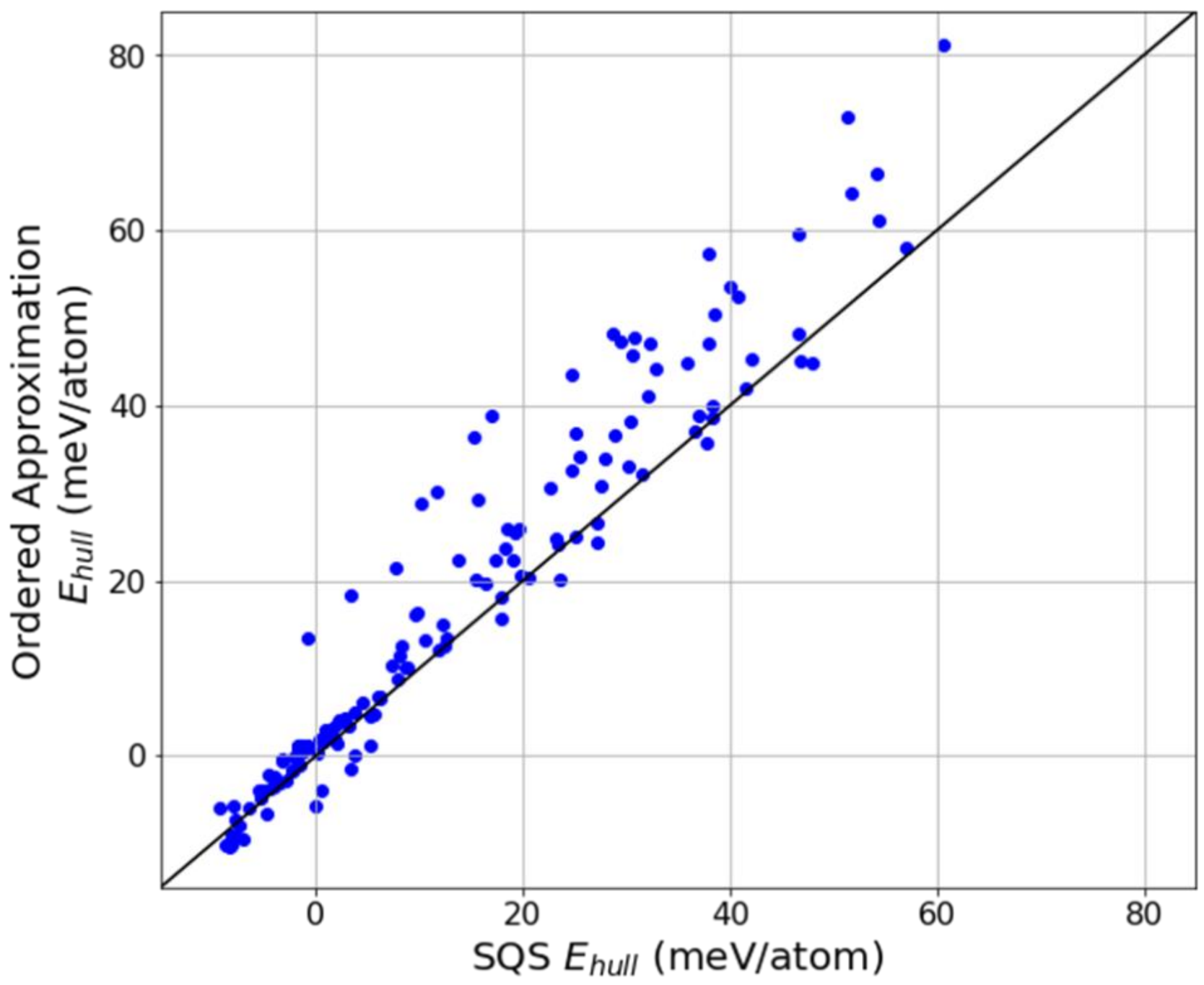

Figure S1 – Comparison of the $E_{hull}$ from SQS and an ordered cell approximation

Figure S1 shows that from the set of SQS computations that were run, the ordered approximation agrees quite well with the energy approximated by the minimum sized supercell, despite not truly representing a disordered compound. In general, the SQS has a slightly lower energy, with an average difference of about 4 meV/atom. The standard deviation of the difference is 6 meV/atom, which is a reasonable level of accuracy for this screening step. The largest discrepancy is in the $Ca(Zn_xCd_{1-x})_2P_2$ for which the ordered cell is 22 meV/atom higher in energy than the SQS. Out of the top 10 systems with the largest differences, nine contain Zn and six of those have Zn-Cd mixing. Our screening criterion to pass to the next step was set to $E_{hull} < 70\ meV/atom$, which is high to account for these errors.

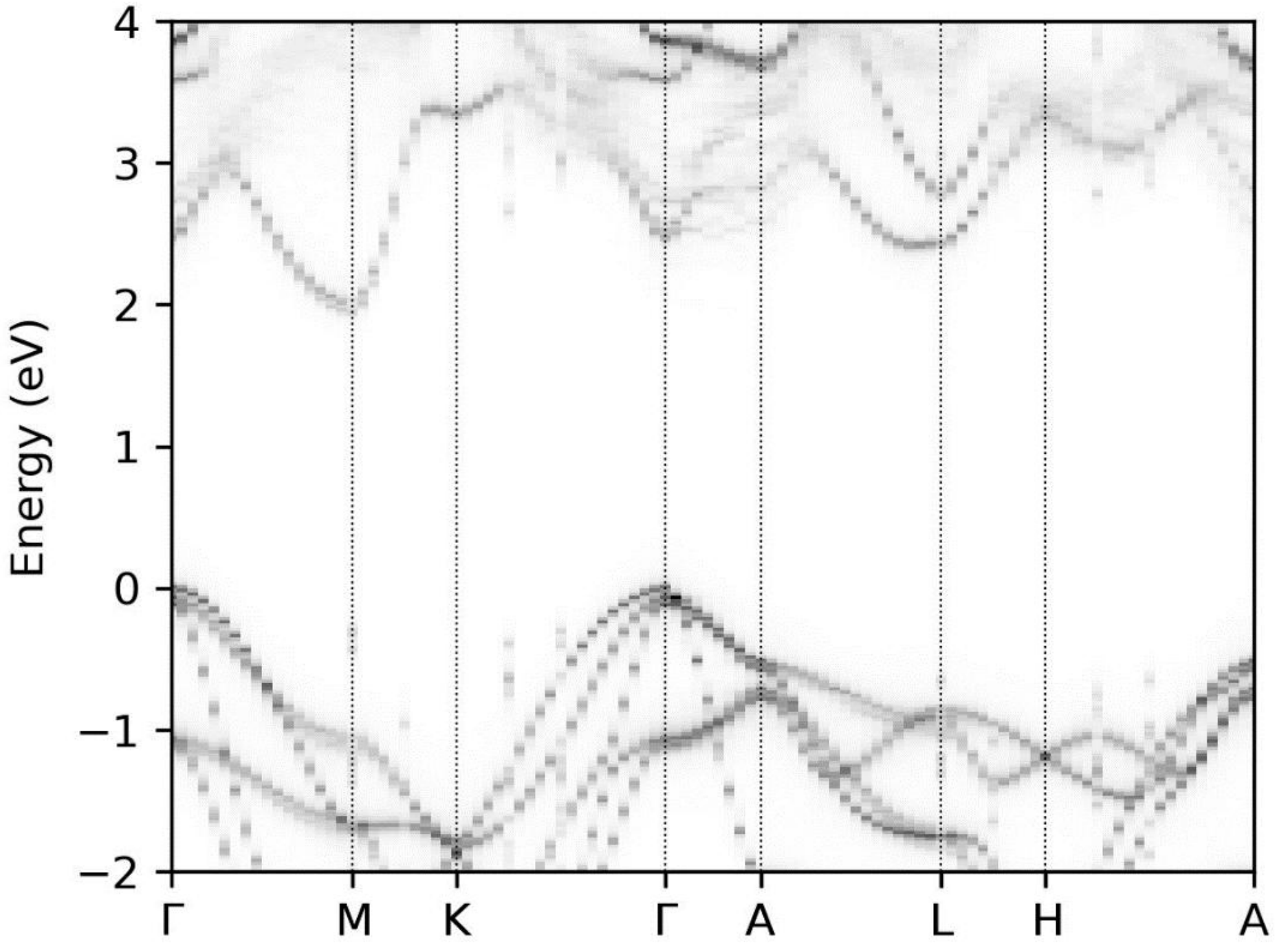

Figure S2 -Unfolded Band structure of $Ca(Cd_{0.25}Mg_{0.75})_2P_2$

Table S3 – Comparison of the unit cell volume and lattice parameters

| Method | Sample | Unit cell volume, V (Å$^3$) | ΔV (Å$^3$) | a (Å) | Δ a (Å) | c (Å) | Δ c (Å) |
|---|---|---|---|---|---|---|---|
| DFT (PBE) | **$CaZn_2P_2$** | 97.202 | - | 4.0497 | - | 6.8432 | - |
| | **$CaMg_2P_2$** | 109.021 | 11.818 | 4.2531 | 0.203 | 6.9593 | 0.116 |
| Experimental (PXRD) | **$CaZn_2P_2$** | 96.523 | - | 4.0372 | - | 6.8382 | - |
| | **$Ca(Zn_{1-x}Mg_x)_2P_2$** | 98.999 | 2.476 | 4.0871 | .0499 | 6.8648 | 0.027 |

Table S4 - Calculated Interaction and Bowing Parameters

| Formula | Ω (eV/atom) | b (eV) |
|---|---|---|
| **$Ba_{1-x}Ca_xCd_2As_2$** | 0.035 | 0.38 |
| **$Ba_{1-x}Ca_xCd_2Bi_2$** | 0.007 | 0 |
| **$Ba_{1-x}Ca_xCd_2N_2$** | 0.102 | 0 |
| **$Ba_{1-x}Ca_xCd_2P_2$** | 0.049 | 0.38 |
| **$Ba_{1-x}Ca_xCd_2Sb_2$** | 0.013 | 0.16 |
| **$Ba_{1-x}Ca_xMg_2As_2$** | 0.065 | 0.54 |
| **$Ba_{1-x}Ca_xMg_2Bi_2$** | 0.029 | 0.34 |
| **$Ba_{1-x}Ca_xMg_2P_2$** | 0.076 | 0.53 |
| **$Ba_{1-x}Ca_xMg_2Sb_2$** | 0.042 | - |
| **$Ba_{1-x}Ca_xZn_2As_2$** | 0.037 | 0.35 |
| **$Ba_{1-x}Ca_xZn_2Bi_2$** | 0.008 | 0 |
| **$Ba_{1-x}Ca_xZn_2P_2$** | 0.056 | 0.65 |
| **$Ba_{1-x}Ca_xZn_2Sb_2$** | 0.009 | 0 |
| **$Ba_{1-x}Ca_xMg_2N_2$** | 0.146 | 1.5 |
| **$BaCd_2(As_{1-x}P_x)_2$** | -0.007 | 0.16 |
| **$BaCd_2(Bi_{1-x}As_x)_2$** | 0.077 | 0.3 |
| **$BaCd_2(Bi_{1-x}P_x)_2$** | 0.154 | 1.36 |
| **$BaCd_2(Bi_{1-x}Sb_x)_2$** | -0.014 | 0 |
| **$BaCd_2(Sb_{1-x}As_x)_2$** | 0.029 | 0.3 |
| **$BaCd_2(Sb_{1-x}P_x)_2$** | 0.09 | 1.33 |
| **$Ba_{1-x}Mg_xCd_2As_2$** | 0.093 | 0.03 |
| **$Ba_{1-x}Mg_xCd_2Bi_2$** | 0.057 | 0 |
| **$Ba_{1-x}Mg_xCd_2N_2$** | 0.096 | 0 |
| **$Ba_{1-x}Mg_xCd_2P_2$** | 0.12 | 1.75 |
| **$Ba_{1-x}Mg_xCd_2Sb_2$** | 0.071 | 0 |
| **$Ba_{1-x}Mg_xZn_2As_2$** | 0.099 | 0.59 |
| **$Ba_{1-x}Mg_xZn_2P_2$** | 0.136 | - |
| **$Ba_{1-x}Mg_xZn_2Sb_2$** | 0.06 | 0 |
| **$BaMg_2(As_{1-x}P_x)_2$** | 0.006 | 0.26 |
| **$BaMg_2(Bi_{1-x}As_x)_2$** | 0.105 | 0.57 |
| **$BaMg_2(Bi_{1-x}P_x)_2$** | 0.187 | 0.75 |
| **$BaMg_2(Bi_{1-x}Sb_x)_2$** | -0.005 | -0.3 |
| **$BaMg_2(Sb_{1-x}As_x)_2$** | 0.062 | 0.5 |
| **$BaMg_2(Sb_{1-x}P_x)_2$** | 0.129 | 0.84 |
| **$Ba(Mg_{1-x}Cd_x)_2As_2$** | -0.032 | -0.19 |
| **$Ba(Mg_{1-x}Cd_x)_2Bi_2$** | -0.02 | 0.82 |
| **$Ba(Mg_{1-x}Cd_x)_2N_2$** | -0.02 | -0.12 |
| **$Ba(Mg_{1-x}Cd_x)_2P_2$** | -0.033 | -0.57 |
| **$Ba(Mg_{1-x}Cd_x)_2Sb_2$** | -0.03 | 0.44 |
| **$Ba(Mg_{1-x}Zn_x)_2As_2$** | 0.006 | 0.01 |
| **$Ba(Mg_{1-x}Zn_x)_2Bi_2$** | 0.02 | 0.82 |
| **$Ba(Mg_{1-x}Zn_x)_2N_2$** | - | -0.93 |
| **$Ba(Mg_{1-x}Zn_x)_2P_2$** | 0.017 | 0.05 |
| **$Ba(Mg_{1-x}Zn_x)_2Sb_2$** | 0.019 | 1.65 |
| **$Ba_{1-x}Sr_xCd_2As_2$** | -0.005 | 0.1 |
| **$Ba_{1-x}Sr_xCd_2Bi_2$** | -0.014 | 0 |
| **$Ba_{1-x}Sr_xCd_2N_2$** | 0.024 | 0 |
| **$Ba_{1-x}Sr_xCd_2P_2$** | 0.003 | 0.17 |
| **$Ba_{1-x}Sr_xCd_2Sb_2$** | -0.016 | 0 |
| **$Ba_{1-x}Sr_xMg_2As_2$** | 0.014 | 0.47 |
| **$Ba_{1-x}Sr_xMg_2Bi_2$** | -0.003 | 0.22 |
| **$Ba_{1-x}Sr_xMg_2P_2$** | 0.018 | 0.45 |
| **$Ba_{1-x}Sr_xMg_2Sb_2$** | 0.005 | 0.41 |
| **$Ba_{1-x}Sr_xZn_2As_2$** | -0.009 | 0.06 |
| **$Ba_{1-x}Sr_xZn_2Bi_2$** | -0.012 | 0 |
| **$Ba_{1-x}Sr_xZn_2N_2$** | - | 0.34 |
| **$Ba_{1-x}Sr_xZn_2P_2$** | 0.001 | 0.35 |
| **$Ba_{1-x}Sr_xZn_2Sb_2$** | -0.019 | 0 |
| **$Ba_{1-x}Sr_xMg_2N_2$** | 0.04 | 0.76 |
| **$BaZn_2(As_{1-x}P_x)_2$** | -0.012 | 0.14 |
| **$BaZn_2(Bi_{1-x}Sb_x)_2$** | -0.012 | 0 |
| **$BaZn_2(Sb_{1-x}As_x)_2$** | 0.039 | - |
| **$Ba(Zn_{1-x}Cd_x)_2As_2$** | 0.035 | 0.15 |
| **$Ba(Zn_{1-x}Cd_x)_2Bi_2$** | - | 0 |
| **$Ba(Zn_{1-x}Cd_x)_2P_2$** | 0.06 | 0.47 |
| **$Ba(Zn_{1-x}Cd_x)_2Sb_2$** | 0.015 | 0 |
| **$CaCd_2(As_{1-x}P_x)_2$** | -0.007 | 0.11 |
| **$CaCd_2(Bi_{1-x}As_x)_2$** | 0.074 | 0.45 |
| **$CaCd_2(Bi_{1-x}P_x)_2$** | 0.16 | 1.66 |
| **$CaCd_2(Bi_{1-x}Sb_x)_2$** | -0.021 | 0.16 |
| **$CaCd_2(Sb_{1-x}As_x)_2$** | 0.033 | 0.59 |
| **$CaCd_2(Sb_{1-x}P_x)_2$** | 0.102 | 1.51 |
| **$Ca_{1-x}Mg_xCd_2As_2$** | -0.01 | 0.33 |
| **$Ca_{1-x}Mg_xCd_2Bi_2$** | -0.012 | 0 |
| **$Ca_{1-x}Mg_xCd_2N_2$** | - | 0 |
| **$Ca_{1-x}Mg_xCd_2P_2$** | 0.003 | 1.03 |
| **$Ca_{1-x}Mg_xCd_2Sb_2$** | -0.012 | 0.16 |
| **$Ca_{1-x}Mg_xZn_2As_2$** | -0.021 | 0.55 |
| **$Ca_{1-x}Mg_xZn_2N_2$** | 0.05 | - |
| **$Ca_{1-x}Mg_xZn_2P_2$** | -0.007 | 0.93 |
| **$Ca_{1-x}Mg_xZn_2Sb_2$** | -0.026 | 0 |
| **$CaMg_2(As_{1-x}P_x)_2$** | 0.005 | 0.37 |
| **$CaMg_2(Bi_{1-x}As_x)_2$** | 0.121 | 0.66 |
| **$CaMg_2(Bi_{1-x}P_x)_2$** | 0.217 | 0.83 |
| **$CaMg_2(Bi_{1-x}Sb_x)_2$** | -0.008 | 0.17 |
| **$CaMg_2(Sb_{1-x}As_x)_2$** | 0.073 | 0.61 |
| **$CaMg_2(Sb_{1-x}P_x)_2$** | 0.152 | 0.78 |
| **$Ca(Mg_{1-x}Cd_x)_2As_2$** | -0.031 | -0.12 |
| **$Ca(Mg_{1-x}Cd_x)_2Bi_2$** | -0.035 | 0.74 |
| **$Ca(Mg_{1-x}Cd_x)_2N_2$** | 0.02 | 2.41 |
| **$Ca(Mg_{1-x}Cd_x)_2P_2$** | -0.034 | -0.6 |
| **$Ca(Mg_{1-x}Cd_x)_2Sb_2$** | -0.033 | 0.23 |
| **$Ca(Mg_{1-x}Zn_x)_2As_2$** | 0 | 0.37 |
| **$Ca(Mg_{1-x}Zn_x)_2Bi_2$** | -0.006 | 0.74 |
| **$Ca(Mg_{1-x}Zn_x)_2N_2$** | -0.057 | 0.31 |
| **$Ca(Mg_{1-x}Zn_x)_2P_2$** | 0.013 | 0.12 |
| **$Ca(Mg_{1-x}Zn_x)_2Sb_2$** | 0 | 1.32 |
| **$CaZn_2(As_{1-x}P_x)_2$** | -0.013 | -0.28 |
| **$CaZn_2(Bi_{1-x}As_x)_2$** | 0.105 | 0.72 |
| **$CaZn_2(Bi_{1-x}P_x)_2$** | 0.216 | 1.71 |
| **$CaZn_2(Bi_{1-x}Sb_x)_2$** | -0.025 | 0 |
| **$CaZn_2(Sb_{1-x}As_x)_2$** | 0.039 | 0.72 |
| **$CaZn_2(Sb_{1-x}P_x)_2$** | 0.129 | 1.71 |
| **$Ca(Zn_{1-x}Cd_x)_2As_2$** | 0.047 | 0.17 |
| **$Ca(Zn_{1-x}Cd_x)_2Bi_2$** | 0.004 | 0 |
| **$Ca(Zn_{1-x}Cd_x)_2N_2$** | 0.109 | 1.25 |
| **$Ca(Zn_{1-x}Cd_x)_2P_2$** | 0.068 | 0.11 |
| **$Ca(Zn_{1-x}Cd_x)_2Sb_2$** | 0.013 | 0.16 |
| **$Mg(Cd_{1-x}Mg_x)_2As_2$** | -0.041 | 1 |
| **$Mg(Cd_{1-x}Mg_x)_2Bi_2$** | -0.045 | 0 |
| **$Mg(Cd_{1-x}Mg_x)_2N_2$** | -0.216 | 3.08 |
| **$Mg(Cd_{1-x}Mg_x)_2P_2$** | -0.036 | 0.31 |
| **$Mg(Cd_{1-x}Mg_x)_2Sb_2$** | -0.049 | 0.33 |
| **$Mg(Zn_{1-x}Mg_x)_2As_2$** | -0.038 | 1.41 |
| **$Mg(Zn_{1-x}Mg_x)_2Bi_2$** | -0.055 | 0 |
| **$Mg(Zn_{1-x}Mg_x)_2N_2$** | - | 0.75 |
| **$Mg(Zn_{1-x}Mg_x)_2P_2$** | -0.019 | 0.68 |
| **$Mg(Zn_{1-x}Mg_x)_2Sb_2$** | -0.039 | 0.33 |
| **$Mg_3(As_{1-x}P_x)_2$** | -0.004 | 0.36 |
| **$Mg_3(Bi_{1-x}As_x)_2$** | 0.09 | 1.72 |
| **$Mg_3(Bi_{1-x}P_x)_2$** | 0.153 | - |
| **$Mg_3(Bi_{1-x}Sb_x)_2$** | -0.031 | 0.33 |
| **$Mg_3(Sb_{1-x}As_x)_2$** | 0.053 | 1.15 |
| **$MgCd_2(As_{1-x}P_x)_2$** | -0.032 | 0.73 |
| **$MgCd_2(Bi_{1-x}Sb_x)_2$** | -0.036 | 0 |
| **$MgCd_2(Sb_{1-x}As_x)_2$** | 0.006 | 0 |
| **$MgCd_2(Sb_{1-x}P_x)_2$** | 0.017 | 0.78 |
| **$MgZn_2(As_{1-x}P_x)_2$** | -0.037 | 0.88 |
| **$MgZn_2(Sb_{1-x}As_x)_2$** | 0.019 | 0.51 |
| **$MgZn_2(Sb_{1-x}P_x)_2$** | 0.108 | - |
| **$Mg(Zn_{1-x}Cd_x)_2As_2$** | 0.02 | 0.51 |
| **$Mg(Zn_{1-x}Cd_x)_2N_2$** | -0.097 | 0.45 |
| **$Mg(Zn_{1-x}Cd_x)_2P_2$** | - | 0.41 |
| **$Mg(Zn_{1-x}Cd_x)_2Sb_2$** | -0.003 | 0 |
| **$Sr_{1-x}Ca_xCd_2As_2$** | -0.01 | 0.21 |
| **$Sr_{1-x}Ca_xCd_2Bi_2$** | -0.022 | 0 |
| **$Sr_{1-x}Ca_xCd_2N_2$** | 0.015 | 0 |
| **$Sr_{1-x}Ca_xCd_2P_2$** | -0.003 | 0.03 |
| **$Sr_{1-x}Ca_xCd_2Sb_2$** | -0.019 | 0.16 |
| **$Sr_{1-x}Ca_xMg_2As_2$** | 0.007 | 0.38 |
| **$Sr_{1-x}Ca_xMg_2Bi_2$** | -0.009 | 0.32 |
| **$Sr_{1-x}Ca_xMg_2P_2$** | 0.011 | 0.12 |
| **$Sr_{1-x}Ca_xMg_2Sb_2$** | 0.001 | 0.46 |
| **$Sr_{1-x}Ca_xZn_2As_2$** | -0.018 | -0.13 |
| **$Sr_{1-x}Ca_xZn_2Bi_2$** | -0.021 | 0 |
| **$Sr_{1-x}Ca_xZn_2N_2$** | 0.025 | 0.1 |
| **$Sr_{1-x}Ca_xZn_2P_2$** | -0.009 | 0.24 |
| **$Sr_{1-x}Ca_xZn_2Sb_2$** | -0.031 | 0 |
| **$Sr_{1-x}Ca_xMg_2N_2$** | 0.031 | 0.06 |
| **$SrCd_2(As_{1-x}P_x)_2$** | -0.007 | 0.02 |
| **$SrCd_2(Bi_{1-x}As_x)_2$** | 0.079 | 0.26 |
| **$SrCd_2(Bi_{1-x}P_x)_2$** | 0.163 | 1.39 |
| **$SrCd_2(Bi_{1-x}Sb_x)_2$** | -0.017 | 0 |
| **$SrCd_2(Sb_{1-x}As_x)_2$** | 0.032 | 0.26 |
| **$SrCd_2(Sb_{1-x}P_x)_2$** | 0.099 | 1.35 |
| **$Sr_{1-x}Mg_xCd_2As_2$** | 0.031 | 0.08 |
| **$Sr_{1-x}Mg_xCd_2Bi_2$** | 0.015 | 0 |
| **$Sr_{1-x}Mg_xCd_2N_2$** | - | 0 |
| **$Sr_{1-x}Mg_xCd_2P_2$** | 0.05 | 1.28 |
| **$Sr_{1-x}Mg_xCd_2Sb_2$** | 0.021 | 0 |
| **$Sr_{1-x}Mg_xZn_2As_2$** | 0.022 | - |
| **$Sr_{1-x}Mg_xZn_2N_2$** | 0.154 | 0.88 |
| **$Sr_{1-x}Mg_xZn_2P_2$** | 0.047 | - |
| **$Sr_{1-x}Mg_xZn_2Sb_2$** | 0.008 | 0 |
| **$SrMg_2(As_{1-x}P_x)_2$** | 0.005 | - |
| **$SrMg_2(Bi_{1-x}As_x)_2$** | 0.116 | - |
| **$SrMg_2(Bi_{1-x}P_x)_2$** | 0.207 | 0.91 |
| **$SrMg_2(Bi_{1-x}Sb_x)_2$** | -0.007 | -0.4 |
| **$SrMg_2(Sb_{1-x}As_x)_2$** | 0.069 | 0.66 |
| **$SrMg_2(Sb_{1-x}P_x)_2$** | 0.144 | 1.1 |
| **$Sr(Mg_{1-x}Cd_x)_2As_2$** | -0.033 | 0.18 |
| **$Sr(Mg_{1-x}Cd_x)_2Bi_2$** | -0.029 | 0.75 |
| **$Sr(Mg_{1-x}Cd_x)_2N_2$** | 0.001 | 1.85 |
| **$Sr(Mg_{1-x}Cd_x)_2P_2$** | -0.034 | -0.76 |
| **$Sr(Mg_{1-x}Cd_x)_2Sb_2$** | -0.032 | 0.39 |
| **$Sr(Mg_{1-x}Zn_x)_2As_2$** | 0.002 | 0.4 |
| **$Sr(Mg_{1-x}Zn_x)_2Bi_2$** | 0.006 | 0.75 |
| **$Sr(Mg_{1-x}Zn_x)_2N_2$** | -0.055 | 0.26 |
| **$Sr(Mg_{1-x}Zn_x)_2P_2$** | 0.015 | 0.28 |
| **$Sr(Mg_{1-x}Zn_x)_2Sb_2$** | 0.008 | 1.54 |
| **$SrZn_2(As_{1-x}P_x)_2$** | -0.013 | 0.25 |
| **$SrZn_2(Bi_{1-x}As_x)_2$** | 0.1 | 0.44 |
| **$SrZn_2(Bi_{1-x}P_x)_2$** | 0.205 | 1.7 |
| **$SrZn_2(Bi_{1-x}Sb_x)_2$** | -0.022 | 0 |
| **$SrZn_2(Sb_{1-x}As_x)_2$** | 0.038 | 0.44 |
| **$SrZn_2(Sb_{1-x}P_x)_2$** | 0.122 | 1.7 |
| **$Sr(Zn_{1-x}Cd_x)_2As_2$** | 0.04 | 0.07 |
| **$Sr(Zn_{1-x}Cd_x)_2Bi_2$** | 0.008 | 0 |
| **$Sr(Zn_{1-x}Cd_x)_2N_2$** | 0.112 | 0.9 |
| **$Sr(Zn_{1-x}Cd_x)_2P_2$** | 0.061 | 0.31 |
| **$Sr(Zn_{1-x}Cd_x)_2Sb_2$** | 0.013 | 0 |

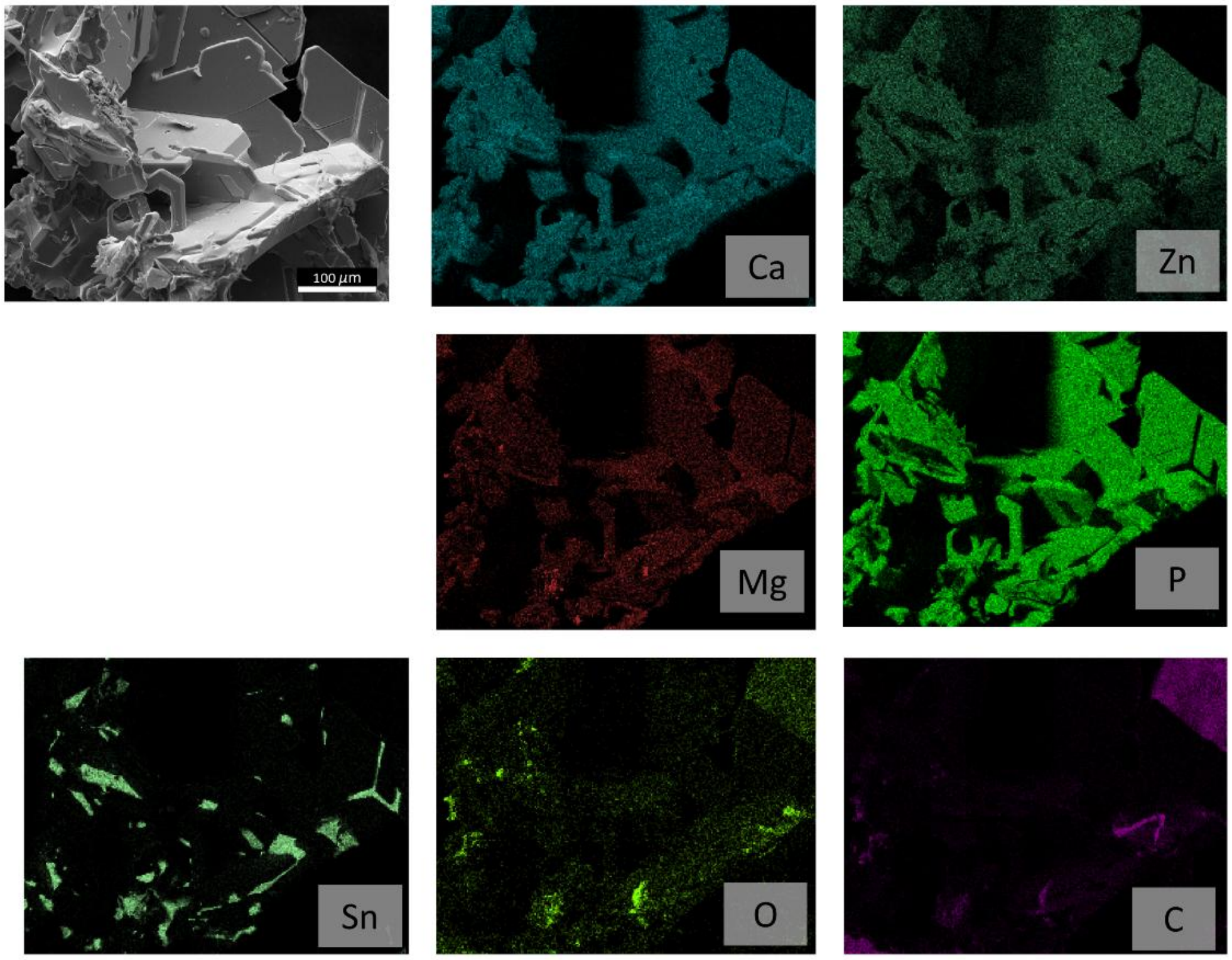

Figure S3 – EDS mapping of the $Ca(Zn_{1-x}Mg_x)_2P_2$ sample for the elements: Ca, Zn, Mg, P, Sn, O, and C. The scale bar is 100 µm.

Table S5 – Elemental composition from EDS. The carbon signal comes from the carbon tape used to support the crystals.

| Element | Atomic % |
|---|---|
| C | 71(7) |
| O | 9(1) |
| Mg | 1.5(1) |
| P | 6.4(3) |
| Ca | 4.0(1) |
| Zn | 7.5(2) |
| Sn | 0.90(3) |

Table S6: Single-crystal X-ray diffraction data of **pristine $CaZn_2P_2$**.

| | **Sample 1** | **Sample 2** | **Sample 3** |
|---|---|---|---|
| Empirical formula | | $CaZn_2P_2$ | |
| Formula weight | | 232.76 | |
| Temperature (K) | | 100(1) | |
| Crystal system | | trigonal | |
| Space group | | $P\bar{3}m1$ | |
| $a$, $b$ (Å) | 4.0352(2) | 4.0340(3) | 4.0349(2) |
| $c$ (Å) | 6.8198(4) | 6.8150(5) | 6.8217(4) |
| Volume (Å$^3$) | 96.17(1) | 96.04(1) | 96.18(1) |
| $Z$ | | 1 | |
| $\rho_{calc}$ (g/cm$^3$) | 4.019 | 4.024 | 4.019 |
| Crystal size (mm$^{-1}$) | 14.38 | 14.40 | 14.38 |
| $F$ (000) | 110.0 | 110.0 | 110.0 |
| Crystal size (mm$^3$) | 0.12 × 0.1 × 0.02 | 0.12 × 0.11 × 0.02 | 0.14 × 0.11 × 0.02 |
| Radiation | | Mo Kα (λ = 0.71073) | |
| $2\boldsymbol{\theta}$ range for data collection (°) | 5.974 to 53.46 | 5.978 to 53.938 | 5.972 to 53.896 |
| Completeness to $2\boldsymbol{\theta}$ ($2\boldsymbol{\theta}$ = 50.484°) | 100% | 100% | 100% |
| Data/restraints/parameters | 99/0/9 | 102/0/9 | 102/0/9 |
| Goodness-of-fit on $F^2$ | 1.07 | 1.09 | 1.11 |
| $R_1$[a] $wR_2$[b] ($\boldsymbol{I} \geq 2\boldsymbol{\sigma}(\boldsymbol{I})$) | $R_1$ = 0.022 $wR_2$ = 0.046 | $R_1$ = 0.024 $wR_2$ = 0.051 | $R_1$ = 0.022 $wR_2$ = 0.048 |
| $R_1$[a] $wR_2$[b] (all data) | $R_1$ = 0.022 $wR_2$ = 0.046 | $R_1$ = 0.024 $wR_2$ = 0.051 | $R_1$ = 0.022 $wR_2$ = 0.048 |
| Largest diff. peak/hole (e Å$^{-3}$) | 0.75 and-0.63 | 0.69 and -1.21 | 0.66 and -0.94 |
| $\boldsymbol{U}_{11}/\boldsymbol{U}_{33}$ for Zn | 0.648 | 0.946 | 0.974 |

[a] $R_1 = \sum||F_o| - |F_c||/\sum|Fo|$. [b] $wR_2 = \left\{\frac{\sum\left[w\left(F_o^2 - F_c^2\right)^2\right]}{\sum\left[w\left(F_o^2\right)^2\right]}\right\}^{1/2}$.

Table S7: Single-crystal X-ray diffraction data of **Mg-alloyed $CaZn_2P_2$**.

| | **Sample 1** | **Sample 2** | **Sample 3** |
|---|---|---|---|
| Empirical formula | $CaZn_{1.36}Mg_{0.64(2)}P_2$ | $CaZn_{1.59}Mg_{0.41(2)}P_2$ | $Ca\ Zn_{1.56}Mg_{0.44(2)}P_2$ |
| Formula weight | 206.48 | 215.93 | 214.69 |
| Temperature (K) | | 100(1) | |
| Crystal system | | trigonal | |
| Space group | | $P\bar{3}m1$ | |
| $a$, $b$ (Å) | 4.0997(5) | 4.0716(2) | 4.0735(2) |
| $c$ (Å) | 6.8581(8) | 6.8465(4) | 6.8508(4) |
| Volume (Å$^3$) | 99.82(3) | 98.295(11) | 98.448(11) |
| $Z$ | | 1 | |
| $\rho_{calc}$ (g/cm$^3$) | 3.435 | 3.648 | 3.621 |
| $\mu$ (mm$^{-1}$) | 10.20 | 11.69 | 11.50 |
| $F$ (000) | 98.0 | 103.0 | 102.0 |
| Crystal size (mm$^3$) | 0.15 × 0.1 × 0.02 | 0.09 × 0.08 × 0.02 | 0.18 × 0.15 × 0.05 |
| Radiation | | Mo Kα (λ = 0.71073) | |
| 2$\boldsymbol{\theta}$ range for data collection (°) | 5.94 to 53.4 | 5.95 to 53.93 | 5.946 to 53.902 |
| Completeness to 2$\boldsymbol{\theta}$ (2$\boldsymbol{\theta}$ = 50.484°) | 100% | 100% | 100% |
| Data/restraints/parameters | 106/0/10 | 105/0/10 | 106/0/11 |
| Goodness-of-fit on $F^2$ | 1.16 | 1.07 | 1.11 |
| $R_1$[a] <br> $wR_2$[b] ($I \geq 2\sigma(I)$) | $R_1$ = 0.022 <br> $wR_2$ = 0.050 | $R_1$ = 0.023 <br> $wR_2$ = 0.050 | $R_1$ = 0.024 <br> $wR_2$ = 0.053 |
| $R_1$[a] <br> $wR_2$[b] (all data) | $R_1$ = 0.025 <br> $wR_2$ = 0.052 | $R_1$ = 0.023 <br> $wR_2$ = 0.050 | $R_1$ = 0.024 <br> $wR_2$ = 0.053 |
| Largest diff. peak/hole (e Å$^{-3}$) | 0.61 and -0.98 | 0.58 and -1.34 | 0.67 and -1.30 |
| $\boldsymbol{U}_{11}$/$\boldsymbol{U}_{33}$ for Zn/Mg | 1.435 | 2.272 | 1.142 |

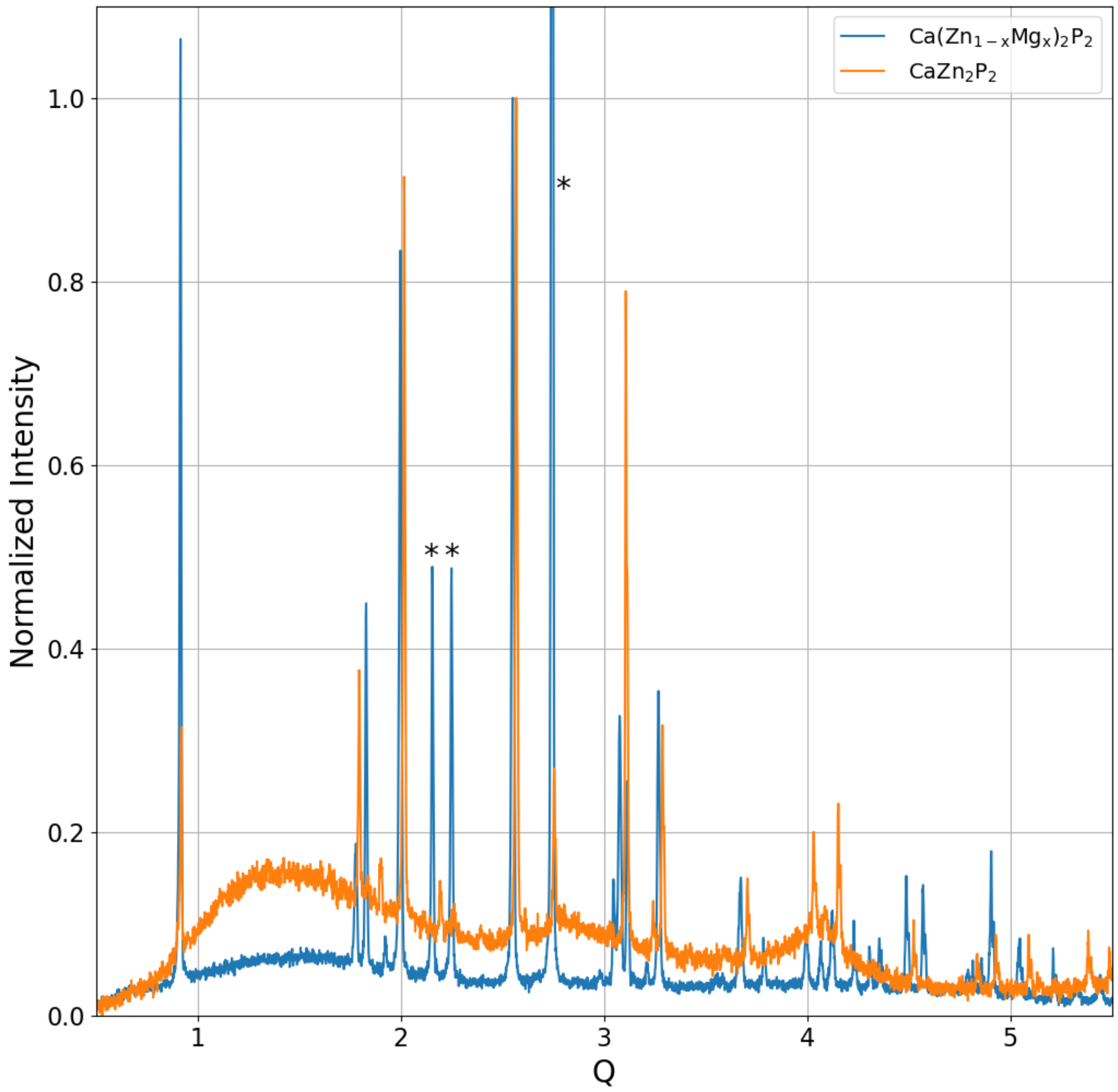

Figure S4 – PXRD patterns of the pristine $CaZn_2P_2$ and Mg-alloyed $CaZn_2P_2$. Intensities are normalized to the peak near Q=2.55. Peaks marked * are due to the presences of Sn impurity.

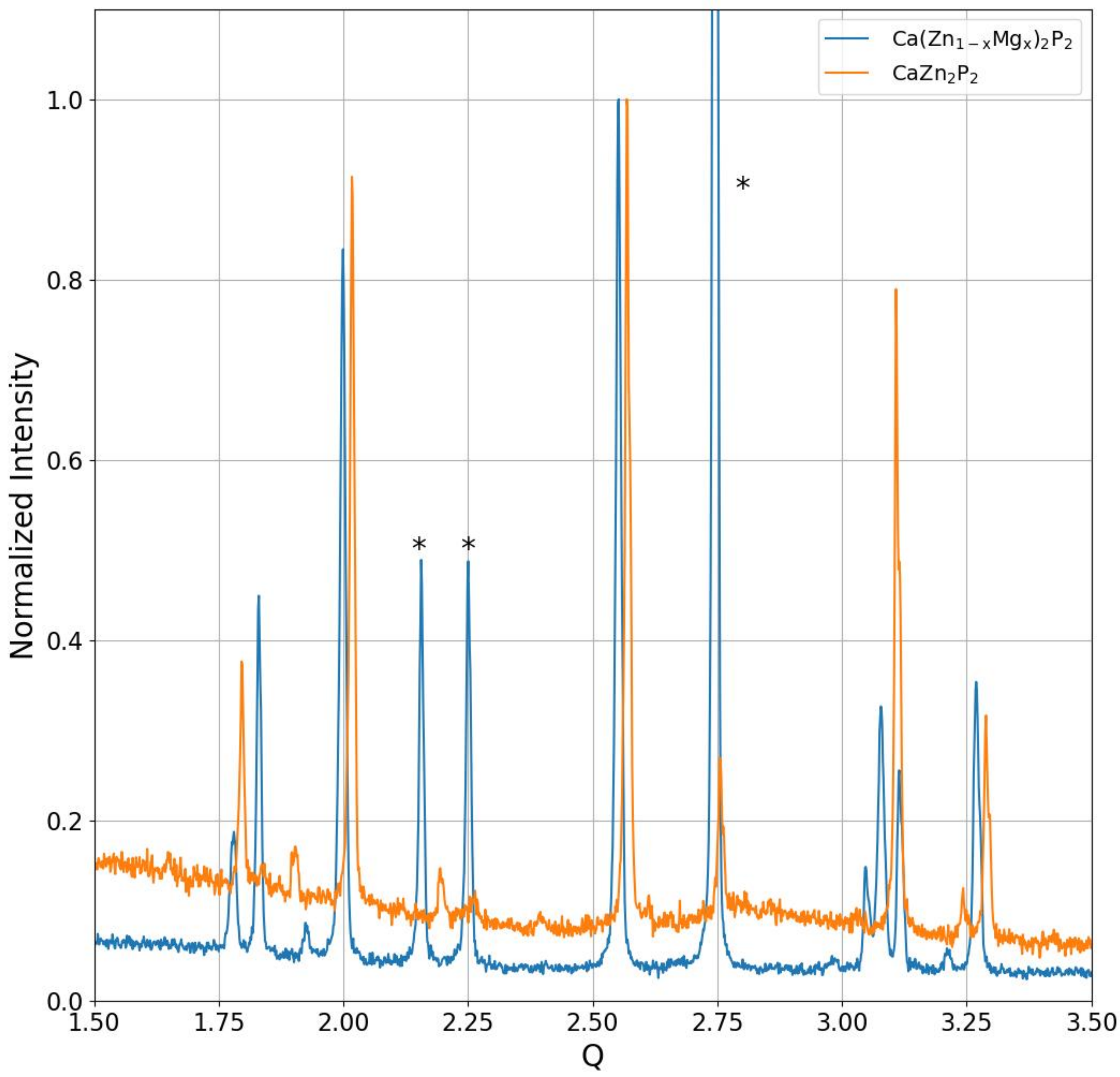

Figure S5 – PXRD patterns of the pristine $CaZn_2P_2$ and Mg-alloyed $CaZn_2P_2$ in the range of $1.5<Q<3.5$. Intensities are normalized to the peak near Q=2.55. Peaks marked * are due to the presence of Sn impurity.

*References*

(1) Imasato, K.; Wood, M.; Anand, S.; Kuo, J. J.; Snyder, G. J. Understanding the High Thermoelectric Performance of Mg3Sb2-Mg3Bi2 Alloys. *Advanced Energy and Sustainability Research* **2022**, *3* (6). https://doi.org/10.1002/aesr.202100208.

(2) Ponnambalam, V.; Morelli, D. T. On the Thermoelectric Properties of Zintl Compounds Mg3Bi2−x Pn x (Pn = P and Sb). *J Electron Mater* **2013**, *42* (7), 1307–1312. https://doi.org/10.1007/s11664-012-2417-7.

(3) Li, A.; Nan, P.; Wang, Y.; Gao, Z.; Zhang, S.; Han, Z.; Zhao, X.; Ge, B.; Fu, C.; Zhu, T. Chemical Stability and Degradation Mechanism of Mg3Sb2-Bi Thermoelectrics towards Room-Temperature Applications. *Acta Mater* **2022**, *239*, 118301. https://doi.org/10.1016/j.actamat.2022.118301.

(4) Imasato, K.; Anand, S.; Gurunathan, R.; Snyder, G. J. The Effect of Mg3As2 Alloying on the Thermoelectric Properties of N-Type Mg3(Sb, Bi)2. *Dalton Transactions* **2021**, *50* (27), 9376–9382. https://doi.org/10.1039/d1dt01600h.

(5) Katsube, R.; Nose, Y. Experimental Investigation of Phase Equilibria around a Ternary Compound Semiconductor Mg(Mg Zn1-)2P2 in the Mg–P–Zn System at 300 °C Using Sn Flux. *J Solid State Chem* **2019**, *280* (April), 120983. https://doi.org/10.1016/j.jssc.2019.120983.

(6) Tortorella, D. S.; Ghosh, K.; Bobev, S. Realization of a Trigonal Mg3–Zn P2 Intermediate Solid Solution between the Binary Cubic Mg3P2 and Tetragonal Zn3P2 End Members. *J Solid State Chem* **2025**, *344* (September 2024), 125184. https://doi.org/10.1016/j.jssc.2025.125184.

(7) Ahmadpour, F.; Kolodiazhnyi, T.; Mozharivskyj, Y. Structural and Physical Properties of Mg3−xZnxSb2 (X=0–1.34). *J Solid State Chem* **2007**, *180* (9), 2420–2428. https://doi.org/10.1016/j.jssc.2007.06.011.

(8) Wu, L.; Zhou, Z.; Han, G.; Zhang, B.; Yu, J.; Wang, H.; Chen, Y.; Lu, X.; Wang, G.; Zhou, X. Realizing High Thermoelectric Performance in P-Type CaZn2Sb2-Alloyed Mg3Sb2-Based Materials via Band and Point Defect Engineering. *Chemical Engineering Journal* **2023**, *475* (September), 145988. https://doi.org/10.1016/j.cej.2023.145988.

(9) Jeong, J.; Shim, D.; Choi, M.-H.; Yunxiu, Z.; Kim, D.-H.; Ok, K. M.; You, T.-S. Golden Ratio of the R+/r- for the Structure-Selectivity in the Thermoelectric BaZn2-XCdxSb2 System. *J Alloys Compd* **2024**, *1002* (June), 175272. https://doi.org/10.1016/j.jallcom.2024.175272.

(10) Wood, M.; Aydemir, U.; Ohno, S.; Snyder, G. J. Observation of Valence Band Crossing: The Thermoelectric Properties of CaZn2Sb2 –CaMg2Sb2 Solid Solution. *J Mater Chem A Mater* **2018**, *6* (20), 9437–9444. https://doi.org/10.1039/C8TA02250J.

(11) Tsuji, M.; Hiramatsu, H.; Hosono, H. Tunable Light Emission through the Range 1.8–3.2 EV and p-Type Conductivity at Room Temperature for Nitride Semiconductors, Ca(Mg1–x Znx)2N2 (X= 0–1). *Inorg Chem* **2019**, *58* (18), 12311–12316. https://doi.org/10.1021/acs.inorgchem.9b01811.

(12) Imasato, K.; Kang, S. D.; Ohno, S.; Snyder, G. J. Band Engineering in Mg3Sb2 by Alloying with Mg3Bi2 for Enhanced Thermoelectric Performance. *Mater Horiz* **2018**, *5* (1), 59–64. https://doi.org/10.1039/c7mh00865a.

(13) Vurgaftman, I.; Meyer, J. R.; Ram-Mohan, L. R. Band Parameters for III–V Compound Semiconductors and Their Alloys. *J Appl Phys* **2001**, *89* (11), 5815–5875. https://doi.org/10.1063/1.1368156.

(14) Nicklas, J. W.; Wilkins, J. W. Accurate Ab Initio Predictions of III–V Direct-Indirect Band Gap Crossovers. *Appl Phys Lett* **2010**, *97* (9). https://doi.org/10.1063/1.3485297.